\documentclass[12pt]{revtex4-1}
\usepackage{color}
\usepackage{amsmath}
\usepackage{amssymb}
\usepackage{graphicx}
\usepackage{esint}

\begin{document}

\title{Analytical Treatment of Coherent Excitation Transfer in FMO Complex}
\author{Pallavi Bhattacharyya}
\email[]{pallavi.iisc@gmail.com}

\affiliation{Department of Inorganic and Physical Chemistry\\
 Indian Institute of Science\\
Bangalore 560012\\
India}
\author{K L Sebastian}
\email[]{kls@ipc.iisc.ernet.in}
\homepage[]{http://ipc.iisc.ernet.in/kls.html}
%\thanks{}
%\altaffiliation{}
\affiliation{Department of Inorganic and Physical Chemistry\\
Indian Institute of Science\\
Bangalore 560012\\
India}

\begin{abstract}
We suggest a new method of studying coherence in finite level systems
coupled to the environment and use it for the Hamiltonian that has
been used to  describe the light-harvesting pigment-protein complex. The method works with the adiabatic
states and transforms the Hamiltonian to a form in which the terms
responsible for decoherence and population relaxation are separated
out. Decoherence is then accounted for non-perturbatively, and population
relaxation using a Markovian master equation. Analytical results can
be obtained for the seven level system and the calculations are
very simple for systems with more levels. We apply the treatment to
the seven level system and the results are in excellent agreement
with the exact numerical results of Nalbach \textit{et al} (P. Nalbach, D. Braun, and M. Thorwart, Physical Review E, 84, 041926 (2011)).
Our approach is able to account for decoherence and population relaxation
separately. It is found that decoherence only causes damping of oscillations,
and does not lead to transfer to the reaction centre. Population relaxation
is necessary for efficient transfer to the reaction centre, in agreement
with earlier findings.  Our results show that the transformation to the  adiabatic basis followed by a Redfield type of approach  leads to results in good agreement with exact simulation.
\end{abstract}
\maketitle

\section{Introduction}

Photosynthesis by plants is one of the most ubiquitous phenomena on
earth. It involves the creation of excitations by absorption of photons
and the transfer of this excitation to the reaction centre where the crucial step in photosynthesis, namely the transfer of an
electron  takes place. This transfer of energy from the absorption site to the reaction site is called light harvesting and is highly efficient with more than 95\% of the absorbed energy being transferred to the reaction centre.  One expects  this transfer  to
be  incoherent classical hopping (F\"{o}rster transfer) from one chromophore
to the next.   The reason for this  is that the chromophores
do not exist isolated. They  are surrounded by a protein scaffold
and solvent molecules. Consequently the excitation which might be
considered as a wave encompassing more than a single site/chromophore
is then continuously measured by the surroundings, viz., the proteins and
the solvent. Therefore the superposition is expected
to be rapidly destroyed (within $\backsim10\ fs$) resulting in the localization
of the excitation at a particular site. The excitation now exhibits
particle-like behavior and the transfer takes place through a series
of independent  hops. However, lately there have been some very interesting
experiments reported by Fleming and coworkers \cite{Brixner:2005rz,Engel:2007od,Read:2007kt}
which observe coherent excitation energy transfer over a considerable
period of time.  It should be noted that there have also been studies in which role of coherence was investigated and found to be important in determining the rate of energy transfer \cite{Leegwater1997,KuehnSundstroem1997,Renger&May1998}.

The FMO complex \cite{Fenna:1975,Blankenship:1997,Blankenship:2003}
or the Fenna-Matthews-Olson complex is found in green sulfur bacteria.
It forms a kind of bridge between the peripheral chlorosome antenna
and the reaction centre and is endowed with the important task of
transferring the excitation energy from the antenna to the reaction
centre. It is essentially a trimer of identical subunits, each comprising
seven bacteriocholophyll a molecules. Recent 2D Electronic spectroscopy
studies observe coherence for as long as $660$ fs at $77$ K \cite{Engel:2007od}
and $300$ fs at $277$ K \cite{Panitchayangkoon:2010lo}. The observations
are very surprising for one would expect rapid decoherence for a system
interacting so profusely with its environment \cite{Prezhdo:1998ij,Gilmore:2006vy,Gilmore:JPhysChemA2008,Nalbach:2011gr,Rebentrost:2009ve,Rebentrost:2009kls,Abramavicius:2010qu,Singh:condmat2012}.
Decoherence refers to the loss of coherence, caused by the quantum
nature of the surroundings and can be avoided only by sufficiently
isolating the system, and by keeping the temperatures low. However,
the above experiments suggest that quantum coherence can be maintained
even in wet and hot physiological systems and this has led to a great
deal of interest and to the emergence of what may be referred to as
Quantum Biology \cite{Ball:2011nl,Lloyd:2011dl}.

Theoretical studies for the same were conducted by Fleming and coworkers.
They formulated a treatment where a quantum dynamical equation is
proposed which considers the reorganization dynamics of the surroundings
non-perturbatively in a hierarchical expansion. These predicted the
survival of coherence at room temperature \cite{Ishizaki:2009sd,Ishizaki:2009zm}
as well. There were also similar conclusions from the theoretical
studies by other groups \cite{Mohseni:2008wm,Caruso:2009pp,Olaya-Castro:2011zn}.
Guzik \textit{et al.} \cite{Mohseni:2008wm} developed continuous
quantum walks in the Liouville space. Their studies suggested that
the interplay between the coherent dynamics of the system and the
dissipative influence of the environment assists in more efficient
transport of the excitation compared to the case where the environment
is considered absent. Independent studies by Plenio \textit{et al.} \cite{Caruso:2009pp}
also suggested quantum transport assisted by noise due to environment.
While these are approximate results,  in a very interesting paper,
exact numerical studies have been performed recently for the FMO complex
by Nalbach \textit{et al.} \cite{Nalbach:2011}. It should also be
mentioned that there have also been some studies which ruled out coherence.
These were atomistic simulations by Kleinekath{\"o}fer \cite{olbrichtheory2011}
$\mathit{et}$ $\mathit{al}$., which sought to calculate the spectral
densities for the FMO trimer. The studies proposed that the electron-phonon
coupling was much stronger than what was considered previously. A
stronger coupling with the environment would imply faster and more
efficient destruction of coherence. Path integral Monte Carlo simulations
by the same group \cite{Kleinekathofer2012} suggested that coherence
would not be retained with the initial excitation put on a specific
chromophore. However it could survive if the initial excitation was
delocalised. The point worth noting here is that all the above methods
employed numerical methods or simulations and were computationally
quite expensive. Also, population relaxation and decoherence resulted
from the same term in the Hamiltonian. We also note that the related
problem of coherence in the spin boson problem has been a subject
of a large amount of literature \cite{WeissBook,LeHur2010,LeHurPRB2010}.

At this point, we emphasize that the definition of decoherence
we follow is the one used by Schlosshauer in his book on decoherence
\cite{SchlosshauerBook}. We quote from the book: ``We consistently
reserve the term \textit{decoherence} to describe the consequences
of (usually in practice irreversible) quantum entanglement with some
environment, in agreement with the historically established meaning
and the vast body of literature on environmental decoherence.....
Thus decoherence should be understood as a distinctly quantum-mechanical
effect with no classical analog.'' As a result, we consider population
relaxation to be distinct from decoherence, unlike many other authors.

In this paper we propose an analytical approach for treating such coherences. The main advantage of the treatment is that the evaluation
is numerically inexpensive, and it can be easily performed for the
FMO protein (seven-level system). Further it can be easily extended to systems
with larger number of states unlike the existing approaches which
are computationally expensive and hence difficult to apply to systems
with large number of states. A comparison of results obtained with
our method with the exact numerical studies by Nalbach \cite{Nalbach:2011}
has also been provided. The agreement between the two is excellent.
Our method uses the adiabatic basis of the Hamiltonian and as we show
below, we can treat the population relaxation and decoherence independently
and hence more efficiently. In Section II, we introduce the generalised
electronic Hamiltonian that is commonly used and transform it to the
adiabatic basis in Section III. Section IV deals with the reduced
density matrix of the system which would give us independent expressions
for population relaxation and decoherence. In Section V, the FMO Hamiltonian
is introduced and analytical expressions are given for the reduced
density matrix elements at a time $t.$ Section VI contains the results
and discussions. While almost all our results are concerned with a
situation in which there are no correlations between the phonon baths
for the different chromophores, in Section VII we give a brief discussion
of how spatial correlations affect the energy transfer. We summarize
our results in Section VIII.

\section{The Hamiltonian}

The Hamiltonian that is commonly used has the minimum number of terms,
that are required to account for the phenomena. The FMO complex has
a finite number ($N=7$) of bacteriochlorophyll a molecules. Each bacteriochlorophyll a can
be in the excited state or the ground state. The separation between
these two is large ($\approx12500\; cm^{-1}$) and the two states
are only radiatively coupled. Environmental tuning makes the excitation
energies on different bacteriochlorophylls different, though by small amounts,
thus helping the flow of energy. The energy values relative to the
lowest element are given by the diagonal elements of the matrix in
Eq. (\ref{matrix}). The dipolar coupling between the bacteriochlorophylls is
responsible for the transfer of excitation from one bacteriochlorophyll to the
next. Its magnitude is typically less than $100\; cm^{-1}$. We describe
the electronic part of the Hamiltonian by
\begin{equation}
H_{el}=\sum_{j}\epsilon_{j}|j\rangle\langle j|+\sum_{i,j}(V_{ij}|i\rangle\langle j|+V_{ji}|j\rangle\langle i|),\label{H_el}
\end{equation}
where $|j\rangle$ with $j=1,2...N$ denotes the state in which the
excitation is on the $j^{th}$ bacteriochlorophyll a molecule. $\epsilon_{j}$
is the energy appropriate for this state. Note that in Eq. (\ref{matrix})
the diagonal matrix element for the excitation having the lowest energy
of value $12210\ cm^{-1}$ is taken as the reference. When the excitation
is on the $j^{th}$ site, it interacts with its surroundings (the nuclear degrees of freedom) and this
is responsible for decoherence and relaxation within these states.
The model used for surroundings is a collection of harmonic oscillators.
On the  $j^{th}$ bacteriochlorophyll site, the coupling is to a set of phonons
represented by
\begin{equation}
H_{ph,j}=\frac{1}{2}\sum_{k}(\frac{\hat{p}_{jk}^{2}}{m_{jk}}+m_{jk}\omega_{jk}^{2}q_{jk}^{2}),
\end{equation}
$q_{jk}$ being the position of the $k^{th}$ harmonic oscillator
associated with the $j^{th}$ site. It has a mass $m_{jk}$ and frequency
$\omega_{jk}$. $p_{jk}$ is the momentum operator. Thus the total
phonon system has the Hamiltonian
\begin{equation}
H_{ph}=\sum_{j}H_{ph,j}.\label{eq:H_ph}
\end{equation}
The presence of the excitation on the $j^{th}$ site causes modification
of the phonon Hamiltonian. The excitation causes a shift in the equilibrium positions of the phonons. This is accounted for by $Q_j$ in equations (\ref{shift}) and  (\ref{Hel-ph}). Eq. (\ref{Hel-ph}) implies that this term may also be thought of as a shift of the energy of the $j^{th}$ excited state by the fluctuations of the phonons.
\begin{equation}
Q_{j}=\sum_{k}m_{jk}\nu_{jk}q_{jk}, \label{shift}
\end{equation}
with
\begin{equation}
H_{el-ph}=\sum_{j}Q_{j}|j\rangle\langle j|. \label{Hel-ph}
\end{equation}
Thus the total Hamiltonian is
\begin{equation}
H=H_{el}+H_{ph}+H_{el-ph}.\label{Hamiltonian}
\end{equation}
Interestingly, with this type of coupling to the phonon system, the
time evolution of the excitation is determined just by the spectral
densities for each site, defined by \cite{WeissBook}
\begin{equation}
J_{j}(\omega)=\sum_{k}\frac{m_{jk}\nu_{jk}^{2}}{2\omega_{jk}}\delta(\omega-\omega_{jk}).
\end{equation}
When the $j^{th}$ site is excited, the excitation would cause a shift
in the equilibrium positions of all the harmonic oscillators for this
site. The displacement of the $k^{th}$ oscillator is $\nu_{jk}/\omega_{jk}^{2}$
costing an energy $\frac{m_{jk}\nu_{jk}^{2}}{2\omega_{jk}^{2}}$.
Sum of this for all the oscillators is denoted by $\lambda_{j}=\frac{1}{2}\sum_{k}\frac{m_{jk}\nu_{jk}^{2}}{\omega_{jk}^{2}}$
and is called the reorganization energy for that site. It may be written
as
\begin{equation}
\lambda_{j}=\int_{0}^{\infty}d\omega J_{j}(\omega)/\omega.
\end{equation}

The Hamiltonian has two competing sets of parameters which affect the
energy transport in opposite ways. The first are the off-diagonal
exciton-exciton coupling $V_{ij}$ responsible for energy transfer
and the other being the exciton-phonon coupling, measured by the reorganisation
energy $\lambda_{j}$. If $|V_{ij}|\ \gg\ \lambda_{j}$, we essentially
have delocalised eigenstates instead of energy being localised to
a particular site and the energy transfer is quantum mechanical and
coherent. One has incoherent, F{\"o}rster like transfer in the opposite limit where  $|V_{ij}|\ \ll\ \lambda_{j}$. The fluctuations of the phonons associated with different sites are likely to be independent, and there is some evidence from simulations \cite{Kleinekathofer2012} showing this to be the case.  However, it has also been suggested that the correlations are important and models in which there are correlations have been investigated.   The correlations can be accounted for by
taking \cite{Sarovar:2011}:
\begin{equation}
\langle Q_{i}(t)Q_{j}(0)\rangle=C_{ij}\langle Q_{i}(t)Q_{i}(0)\rangle\label{CMatrix}.
\end{equation}
In the above equation, $C$ is the correlation matrix with the matrix
element $C_{ij}$ denoting the correlation among the chromophores
i and j. Obviously, $C_{ii}=1$. In most of the following, except
in Section VIII, we have assumed that $C_{ij}=\delta_{ij}$ and $J_{j}(\omega)=J(\omega)$
for all $j$.

However, if $\lambda_{j}\ \gg\ |V_{ij}|$, the strong interaction
with the environment makes it difficult for a delocalised state to
survive. Consequently one would expect classical incoherent particle-like
hopping of the excitation from one site to the other. There would
also be an intermediate regime where $|V_{ij}|$ and $\lambda_{j}$
are comparable, and this is the case of the FMO complex.

In the usual approaches, for the intermediate regime conditions, one
would take $H_{el-ph}$ as a perturbation \cite{Ishizaki:2009mw}
and then derive a master equation for the time development of the
reduced density matrix, which then is solved approximately using different
techniques. It is important to realize that in this type of approach,
the same term ($H_{el-ph})$ causes both decoherence and population
relaxation. One would then resort to some approximate way of handling
the perturbation, involving truncating at some order in $Q_{j}$s.
Our approach here follows a different route. We use an approach usual
in the theory of non-adiabaticity effects in chemical reactions \cite{Baer-book}.
We re-express the Hamiltonian in terms of the adiabatic states. In
such a representation, accounting for decoherence becomes easy as
a part of the adiabatic Hamiltonian is responsible for major component
of the decoherence and the non-adiabatic coupling causes population
relaxation. Thus the two important effects of $H_{el-ph}$ viz., decoherence
and population relaxation are separated out (at the lowest order)
and can be accounted for separately, in a natural fashion. Once this
separation is done, the calculation proceeds in a fashion similar
to references \cite{Zhang:1998ri,Yang:2002ss}. In the next section,
we give an outline of the approach.\\

\section{The adiabatic basis and the mapping $T(\mathbf{Q})$}

It is convenient to use the notation $\mathbf{Q}=(Q_{1},Q_{2}...)$.
The survival of coherences for long times implies that there is considerable
delocalization even in presence of the environment. Therefore, it
is natural to expect that the adiabatic eigenfunctions of the Hamiltonian
\begin{equation}
H_{ad}(\mathbf{Q})=H_{el}+H_{el-ph},\label{eq:H_ad}
\end{equation}
would be the best starting point to describe dynamics. Therefore,
we wish to express the total Hamiltonian of Eq. (\ref{Hamiltonian})
in terms of the eigenfunctions $|m(\mathbf{Q})\rangle$, with $m=1,2....N$.
They obey the equation
\begin{equation}
H_{ad}(\mathbf{Q})|m(\mathbf{Q})\rangle=\varepsilon_{m}(\mathbf{Q})|m(\mathbf{Q})\rangle.
\end{equation}
$|m(\mathbf{Q})\rangle$ are obviously linear combinations of $|j\rangle$
with the coefficients dependent on $\mathbf{Q}$. It is also convenient
to introduce creation and annihilation operators for these states
as $c_{m}^{\dagger}(\mathbf{Q})$ and $c_{m}(\mathbf{Q})$ (the operators
have the anticommutator $\{c_{m}^{\dagger}(\mathbf{Q}),c_{n}(\mathbf{Q})\}=\delta_{mn}$).
We can then write the total Hamiltonian as
\begin{equation}
H=\sum_{m}\varepsilon_{m}(\mathbf{Q})c_{m}^{\dagger}(\mathbf{Q})c_{m}(\mathbf{Q})+H_{ph}.\label{H_adb}
\end{equation}
This Hamiltonian has the problem that $c_{m}^{\dagger}(\mathbf{Q})$
does not commute with $H_{ph}$. It would be better if the Hamiltonian
is expressed in terms of $c_{m}^{\dagger}(\mathbf{Q}=0)$ as they
will commute with $H_{ph}$. Note that $\mathbf{Q}=\mathbf{0}$ is
the equilibrium value of $\mathbf{Q}$. Hence we introduce a unitary
transformation which maps $|m(\mathbf{Q})\rangle$ to $|m(\mathbf{Q=0})\rangle$
as $|m(\mathbf{Q})\rangle=T(\mathbf{Q})|m>$. To simplify the appearence
of the equations, we use the notation $|m\rangle$ for $|m(\mathbf{Q}=\mathbf{0})>$
and write the corresponding operator as $c_{m}^{\dagger}$. Obviously,
$c_{m}^{\dagger}=T^{\dagger}(\mathbf{Q})c_{m}^{\dagger}(\mathbf{Q})T(\mathbf{Q})$.
The derivative $\nabla_{\mathbf{Q}}|m(\mathbf{Q})>$ can be
written as $\sum{}_{n}|n(\mathbf{Q})\rangle\langle n(\mathbf{Q})|\nabla_{\mathbf{Q}}|m(\mathbf{Q})>$
and hence we have
\begin{equation}
-i\hbar\nabla_{\mathbf{Q}}c_{m}^{\dagger}(\mathbf{Q})=\sum{}_{n}c_{n}^{\dagger}(\mathbf{Q})\langle n(\mathbf{Q})|(-i\hbar\nabla_{\mathbf{Q}})|m(\mathbf{Q})\rangle.
\end{equation}
From the above, we get \cite{Baer-book}
\begin{equation}
-i\hbar\nabla_{\mathbf{Q}}T(\mathbf{Q})=T(\mathbf{Q})\mathbf{\widehat{A}}(\mathbf{Q}).
\end{equation}
$\widehat{\mathbf{A}}(\mathbf{Q})$ is a vector of dimension $N$
whose $j^{th}$component is the operator $\widehat{A}{}^{j}(\mathbf{Q})$
which may be written as
\begin{equation}
\widehat{A}^{j}(\mathbf{Q})=\sum_{n,m}A_{nm}^{j}(\mathbf{Q})c_{n}^{\dagger}c_{m},
\end{equation}
where $A_{nm}^{j}(\mathbf{Q})=-i\hbar\langle n(\mathbf{Q})|\frac{\partial}{\partial Q_{j}}m(\mathbf{Q})\rangle$.
The Hellmann-Feynman theorem can be used to evaluate $A_{nm}^{j}(\mathbf{Q})$.
For this one uses the result that $\left[\frac{\partial}{\partial Q_{j}},H\right]=\left|j\left\rangle \right\langle j\right|,$
and takes the matrix element of this in the basis of functions $|m(\mathbf{Q})>$
to get $A_{nm}^{j}(\mathbf{Q})=-i\hbar\frac{\langle n(\mathbf{Q})|j\rangle\langle j|m(\mathbf{Q})\rangle}{\varepsilon_{n}(\mathbf{Q)-\varepsilon_{m}(\mathbf{Q)}}}$.
In the following, we shall use the symbols $a,b,c,i$ and $j$ for orbitals on the sites while $m,n,r,r',s'$ and $s$ for
adiabatic eigenstates evaluated at $\mathbf{Q=0}.$ $k$ stands for
the $k^{th}$ harmonic oscillator mode. We now find the transformed
Hamiltonian $\overline{H}=T^{\dagger}(\mathbf{Q})HT(\mathbf{Q})$.
For this we use the equation
\begin{equation}
T^{\dagger}(\mathbf{Q})(\hat{p}_{jk})T(\mathbf{Q})=\hat{p}_{jk}+\frac{\partial Q_{j}}{\partial q_{jk}}A^{j}(\mathbf{Q}).
\end{equation}
Then
\begin{equation}
\overline{H}=\sum_{m}\varepsilon_{m}(\mathbf{Q})c_{m}^{\dagger}c_{m}+\frac{1}{2}\sum_{j,k}\frac{1}{m_{jk}}\left[\hat{p}_{jk}+\sum_{n,m}m_{jk}\nu_{jk}\widehat{A}^{j}(\mathbf{Q})\right]^{2}+\frac{1}{2}\sum_{j,k}m_{jk}\omega_{jk}^{2}q_{jk}^{2}.\label{eq:transformmomentum}
\end{equation}
In the above, the terms linear in $\widehat{A}^{j}(\mathbf{Q})$ are
the non-adiabatic coupling terms. The retention of coherence in the
system implies that these non-adiabatic coupling terms are small.
Hence one would expect the terms quadratic in $\widehat{A}^{j}(\mathbf{Q})$
to be small and we shall neglect them. The Hamiltonian thus becomes
\begin{equation}
\overline{H}=\overline{H}_{0}+\overline{H}_{na}\label{eq:Happroximate-1}
\end{equation}

\begin{equation}
\overline{H}_{0}=\sum_{m}\varepsilon_{m}(\mathbf{Q})c_{m}^{\dagger}c_{m}+\frac{1}{2}\sum_{j,k}\left\{ \frac{\widehat{p}_{jk}^{2}}{m_{jk}}+m_{jk}\omega_{jk}^{2}q_{jk}^{2}\right\} \label{eq:Hzerobar}
\end{equation}
and
\begin{equation}
\overline{H}_{na}=\frac{1}{2}\sum_{j}\left\{ \widehat{P}_{j}\widehat{A}^{j}(\mathbf{Q})+\widehat{A}^{j}(\mathbf{Q})\widehat{P}_{j}\right\} ,
\end{equation}
where
\begin{equation}
\widehat{P}_{j}=\sum_{k}\nu_{jk}\widehat{p}{}_{jk}.\label{eq:P_j
defined}
\end{equation}
With this, all the terms in the Hamiltonian except $\overline{H}_{na}$
are diagonal in the electronic basis consisting of $|m\rangle$. In
$\overline{H}$$,$ the most important term is $\sum_{m}\varepsilon_{m}(\mathbf{Q})c_{m}^{\dagger}c_{m}$.
This term is diagonal in the basis $\{|m\rangle\}$, but note that
the $\mathbf{Q}$ dependence of $\varepsilon_{m}(\mathbf{Q})$ implies
coupling to the surroundings. It would not cause jumps between different
$|m\rangle$ states, but would cause the system to get entangled with
the phonon bath, leading to decoherence. $\overline{H}_{na}$ represents
non-adiabatic coupling and causes transitions between the different
$|m\rangle$ and is responsible for relaxation of the populations
of the adiabatic states $|m\rangle$.

\section{The Reduced Density Matrix}

Let us use the notation $|\alpha>,$ $|\beta>,$ $|\gamma>$ to denote
arbitrary states of the system (we will specify them later - typically,
they would be states in which the excitation is on the sites $a,b$
and $c$).  Let us say we start with an initial
density operator $|\alpha><\alpha|$ $\rho_{ph}(0)$, where \
$\rho_{ph}(0)$ denotes the initial density operator for the phononic
part which is taken to be in equilibrium at a temperature $T$. We
allow the system to evolve in time during the interval $[0,t]$ to
get the total density operator $\rho(t)$ and then find the contribution
of the coherence $|\gamma><\beta|$ to the reduced density matrix
by calculating $\rho_{\beta\gamma}(t)=Tr_{ph}\{|\gamma><\beta|\rho(t)\}$.
$\rho_{\beta\gamma}(t)$ may be written as
\begin{equation}
\rho_{\beta\gamma}(t)=Tr\{c_{\beta}e^{-iHt/\hbar}c_{\alpha}^{\dagger}|0><0|\rho_{ph}(0)c_{\alpha}\ e^{iHt/\hslash}c_{\gamma}^{\dagger}\}.
\end{equation}
In the above, $|0>$ denotes the vacuum state. Introducing identity
as $T(\mathbf{Q})T^{\dagger}(\mathbf{Q})$ and using $H=T(\mathbf{Q})\overline{H}T^{\dagger}(\mathbf{Q})$,
one can write $\rho_{\beta\gamma}(t)$ as: \begin{widetext}
\begin{eqnarray}
\rho_{\beta\gamma}(t)= & Tr_{ph}\{c_{\beta}T(\mathbf{Q})T^{\dagger}(\mathbf{Q})e^{-iHt/\hbar}T(\mathbf{Q})T^{\dagger}(\mathbf{Q})c_{\alpha}^{\dagger}|0><0|\rho_{ph}(0)c_{\alpha}T(\mathbf{Q})T^{\dagger}(\mathbf{Q})\ e^{iHt/\hslash}T(\mathbf{Q})T^{\dagger}(\mathbf{Q})c_{\gamma}^{\dagger}\}\nonumber \\
= & Tr_{ph}\{c_{\beta}T(\mathbf{Q})e^{-i\overline{H}t/\hbar}T^{\dagger}(\mathbf{Q})c_{\alpha}^{\dagger}|0><0|\rho_{ph}(0)c_{\alpha}T(\mathbf{Q})\ e^{i\overline{H}t/\hslash}T^{\dagger}(\mathbf{Q})c_{\gamma}^{\dagger}\}\label{eq:rhobetagamma}
\end{eqnarray}
\end{widetext} We shall denote $T^{\dagger}(\mathbf{Q})c_{\alpha}^{\dagger}|0\rangle\mbox{ as }|\overline{\alpha}\rangle$.
Our calculations are simple if $|\overline{\alpha}\rangle$ is taken
to be an orbital in which the bacteriochlorophyll a on the $a^{th}$ site is excited.
(The alternate choice would be to take $c_{\alpha}^{\dagger}|0\rangle$
to be an excited state on site $a$, but this would lead to slightly
more complicated expressions, without adding to the physics of the
problem). Thus in the following we shall take $|\overline{\alpha}\rangle=|a\rangle,$
$|\overline{\beta}\rangle=|b\rangle$ and $|\overline{\gamma}\rangle=|c\rangle$,
as this simplifies the calculations.

\subsection{Decoherence}

We now split the Hamiltonian in Eq. (\ref{eq:Happroximate-1}) into
the unperturbed part $\overline{H}_{0}$ and the perturbation $\overline{H}_{na}$.
\ Time evolution under $\overline{H}_{0}$ causes the state $|m\rangle$
to \textit{remain in that state only}, even though its energy fluctuates
following variations in $\mathbf{\mathbf{Q}}$, thus leading to decoherence.
Note that time evolution under this Hamiltonian alone will not lead
to any change in the population of the states $|m\rangle$. On the
other hand $\overline{H}_{na}$ causes jumps between the states $|m\rangle$,
and \textit{can lead to changes in the populations}. It is convenient
to use the interaction picture with $\overline{H}_{0}$ as the unperturbed
Hamiltonian. We start with
\begin{equation}
\rho_{bc}(t)=Tr_{ph}\{\langle b|e^{-i\overline{H}t/\hslash}|a\rangle\rho_{ph}(0)\langle a|e^{i\overline{H}t/\hslash}|c\rangle\},\
\end{equation}
and rewrite it as
\begin{align}
\rho_{bc}(t) & =\sum_{m,n,m',n'}Tr_{ph}\{\langle b|m\rangle\langle m|e^{-i\overline{H}t/\hslash}e^{+i\overline{H_{0}}t/\hslash}e^{-i\overline{H}_{0}t/\hslash}|n\rangle\langle n|a\rangle\rho_{ph}(0)\nonumber \\
 & \times\langle a|n'\rangle\langle n'|e^{i\overline{H}_{0}t/\hslash}e^{-i\overline{H_{0}}t/\hslash}e^{i\overline{H}t/\hslash}|m'\rangle\langle m'|c\rangle\}.\label{eq:rhobc}
\end{align}
Defining the time evolution operator in the interaction picture $U_{I}(t)$
by $U_{I}(t)=e^{-i\overline{H}t/\hslash}e^{+i\overline{H}_{0}t/\hslash}$,
we get
\begin{equation}
\rho_{bc}(t)=\sum_{m,n,m',n'}\langle b|m\rangle\langle n|a\rangle\langle a|n'\rangle\langle m'|c\rangle S_{mn,n'm'}(t),\label{eq:rhobc-1-1}
\end{equation}
where
\begin{equation}
S_{mn,n'm'}(t)=Tr_{ph}\{\langle m|U_{I}(t)e^{-i\overline{H}_{0}t/\hslash}|n\rangle\rho_{ph}(0)\langle n'|e^{i\overline{H}_{0}t/\hslash}U_{I}^{\dagger}(t)|m'\rangle\}.\label{eq:DefinitionofS}
\end{equation}
We use the fact that $\overline{H}_{0}$ is diagonal in the adiabatic
basis $|n\rangle$. It is convenient to represent $e^{-i\overline{H}_{0}t/\hbar}|n\rangle$
as $\left(\hat{T}e^{-i\int_{0}^{t}dt_{1}\varepsilon_{n}(\mathbf{Q}(t_{1}))dt_{1}/\hbar}\right)e^{-iH_{ph}t/\hbar}|n\rangle,$
where $\mathbf{Q}(t_{1})=e^{-H_{ph}t_{1}/\hbar}\mathbf{Q}e^{iH_{ph}t_{1}/\hbar}$
and $\hat{T}$ is the time ordering operator. Then
\begin{equation}
S_{mn,n'm'}(t)=Tr_{ph}\{\langle m|U_{I}(t)|n\rangle\left(\hat{T}e^{-i\int_{0}^{t}dt_{1}\varepsilon_{n}(\mathbf{Q}(t_{1}))dt_{1}/\hbar}\right)\rho_{ph}(0)\left(\hat{T}^{\dagger}\, e^{i\int_{0}^{t}dt_{1}\varepsilon_{n'}(\mathbf{Q}(t_{1}))dt_{1}/\hbar}\right)\langle n'|U_{I}^{\dagger}(t)|m'\rangle\}.
\end{equation}
which we can write as
\begin{align}
S_{mn,n'm'}(t) & =Tr_{ph}\{\langle m|U_{I}(t)|n\rangle\rho_{ph}(0)\langle n|U_{I}^{\dagger}(t)|m'\rangle\delta_{nn'}\}\nonumber \\
 & +(1-\delta_{nn'})Tr_{ph}\{\langle m|U_{I}(t)|n\rangle\left(\hat{T}e^{-i\int_{0}^{t}dt_{1}\varepsilon_{n}(\mathbf{Q}(t_{1}))dt_{1}/\hbar}\right)\rho_{ph}(0)\nonumber \\
 & \times\left(\hat{T}^{\dagger}\, e^{i\int_{0}^{t}dt_{1}\varepsilon_{n'}(\mathbf{Q}(t_{1}))dt_{1}/\hbar}\right)\langle n'|U_{I}^{\dagger}(t)|m'\rangle\}
\end{align}
The term $\left(\hat{T}e^{-i\int_{0}^{t}dt_{1}\varepsilon_{n}(\mathbf{Q}(t_{1}))dt_{1}/\hbar}\right)\rho_{ph}(0)\left(\hat{T}^{\dagger}\, e^{i\int_{0}^{t}dt_{1}\varepsilon_{n'}(\mathbf{Q}(t_{1}))dt_{1}/\hbar}\right)$
in the above equation is responsible for decoherence. To proceed further,
we assume that the correlation between decoherence and relaxation
can be neglected and approximate
\begin{align*}
Tr_{ph}\{\langle m|U_{I}(t)|n\rangle\left(\hat{T}e^{-i\int_{0}^{t}dt_{1}\varepsilon_{n}(\mathbf{Q}(t_{1}))dt_{1}/\hbar}\right)\rho_{ph}(0)\left(\hat{T}^{\dagger}\, e^{i\int_{0}^{t}dt_{1}\varepsilon_{n}'(\mathbf{Q}(t_{1}))dt_{1}/\hbar}\right)\langle n'|U_{I}^{\dagger}(t)|m'\rangle\} & \approx\\
Tr_{ph}\{\langle m|U_{I}(t)|n\rangle\rho_{ph}(0)\langle n'|U_{I}^{\dagger}(t)|m'\rangle\}Tr_{ph}\{\left(\hat{T}e^{-i\int_{0}^{t}dt_{1}\varepsilon_{n}(\mathbf{Q}(t_{1}))dt_{1}/\hbar}\right)\rho_{ph}(0)\left(\hat{T}^{\dagger}\, e^{i\int_{0}^{t}dt_{1}\varepsilon_{n}'(\mathbf{Q}(t_{1}))dt_{1}/\hbar}\right)\}.
\end{align*}
Using this in Eq. (\ref{eq:rhobc-1-1}) and Eq. (\ref{eq:DefinitionofS})
we get
\begin{multline}
\rho_{bc}(t)\approx\sum_{m,n,m'}\langle b|m\rangle\langle n|a\rangle\langle a|n\rangle\langle m'|c\rangle p_{mm',nn}(t)+\sum_{m,m',n'\neq n}\langle b|m\rangle\langle n|a\rangle\langle a|n'\rangle\langle m'|c\rangle p_{mm',nn'}(t)D_{nn'}(t).
\end{multline}
We have used the notations $p_{mm',nn'}(t)=Tr_{ph}\{\langle m|U_{I}(t)|n\rangle\rho_{ph}(0)\langle n'|U_{I}^{\dagger}(t)|m'\rangle\}$
and $D_{nn'}(t)=Tr_{ph}\{\left(\widehat{T}^{\dagger}e^{i\int_{0}^{t}\varepsilon_{n'}(\mathbf{Q(}t_{1}))dt_{1}/\hbar}\right)\rho_{ph}(0)\left(\widehat{T}e^{-i\int_{0}^{t}\varepsilon_{n}(\mathbf{Q(}t_{2}))dt_{2}/\hbar}\right)\}$.

\subsection{The time evolution operator $U_{I}(t)$}

$U_{I}(t)$, being the unitary operator in the interaction picture,
obeys the equation $i\hbar\frac{\partial U_{I}(t)}{\partial t}=U_{I}(t)\overline{H}_{na}(t)$
with $\overline{H}_{na}(t)=e^{i\overline{H}_{0}t/\hbar}\overline{H}_{na}e^{-i\overline{H}_{0}t/\hbar}$.
As the system is close to its adiabatic limit, we can take $\overline{H}_{na}$
as a perturbation, approximate it by its value at $\mathbf{Q=0}$,
so that we have
\begin{align}
\overline{H}_{na} & \approx\overline{H}_{na,\mathbf{0}}=\frac{1}{2}\sum_{j}\left\{ \widehat{P}_{j}\widehat{A}^{j}(\mathbf{0})+\widehat{A}^{j}(\mathbf{0})\widehat{P}_{j}\right\} =\sum\widehat{P}_{j}\widehat{A}^{j}(\mathbf{0})\nonumber \\
= & \sum_{j,n,m}\widehat{P}_{j}A_{nm\mathbf{,0}}^{j}\; c_{n}^{\dagger}c_{m}. \label{DefinitionHna}
\end{align}
We now approximate the Hamiltonian in Eq. (\ref{eq:Happroximate-1})
by
\begin{equation}
\overline{H}\simeq\overline{H}_{0}+\overline{H}_{na,\mathbf{0}}.\label{eq:HforMasterEquation}
\end{equation}
Our interest is in the evaluation of $p_{mm',nn'}(t)$. For this,
we introduce $\rho(t)=e^{-i\overline{H}t/\hbar}\rho(0)e^{i\overline{H}t/\hbar}$
and define $\widetilde{\rho}(t)=e^{i\overline{H}_{0}t/\hbar}\rho(t)e^{-i\overline{H}_{0}t/\hbar}$.
Then we note that
\begin{equation}
p_{mm',nn'}(t)=Tr_{ph}\{\langle m|U_{I}(t)|n\rangle\rho_{ph}(0)\langle n'|U_{I}^{\dagger}(t)|m'\rangle\}
\end{equation}
if $\rho(0)$ is taken to be $\rho_{ph}(0)|n\rangle\langle n'|$. {    It is convenient to define an operator $\widetilde{\rho}_{nn'}(t)$ by 
\begin{equation}
\langle m|\widetilde{\rho}_{  nn'}(t)|m'\rangle=
 p_{mm',nn'}(t).
\end{equation}} 
We now proceed to derive an equation for $\langle m|\widetilde{\rho}_{  nn' }(t)|m'\rangle$
for an arbitrary initial density matrix of the form $\rho_{ph}(0)\rho_{e}$
where $\rho_{e}$ is the intial density operator for the electronic
state. $\widetilde{\rho}_{nn'}(t)$ obeys the Liouville equation
\begin{equation}
\frac{\partial\widetilde{\rho}_{  nn' }(t)}{\partial t}=-\frac{i}{\hbar}\left[\overline{H}_{na,\mathbf{0}},\widetilde{\rho}_{  nn' }(t)\right]=L_{na}\left(t\right)\widetilde{\rho}_{  nn' }(t),\label{eq:liouville}
\end{equation}
where $\overline{H}_{na,\mathbf{0}}$ is defined in Eq. (\ref{DefinitionHna}) and $L_{na}(t)$
is defined by the above equation.  {   This equation has to be solved, subject to the initial condition $\rho_{nn'}(0)=|n\rangle \langle n'|$.   Looking at this equation, one realizes that this just the problem of time evolution of initial coherences and populations under the influence of the perturbation $\overline{H}_{na,\mathbf{0}}$.    Also, note that the coherence $|n\rangle \langle n'|$ enters only through the initial condition and that the equation itself is the same for all $|n\rangle \langle n'|$.  }     Following well known methods \cite{Leegwater1997,KuehnSundstroem1997,Zhang:1998ri,Yang:2002ss,Renger&May1998,Renger&MarcusJPCA2003,RengerMarcusJCP2003}, we derive the master equation for the reduced density matrix $\rho_1(t)$(see Appendix \ref{AppendixB} for more details)
\begin{multline}
\frac{\partial\widetilde{\rho}_{1}(t)}{\partial t}=-\sum_{r,r',s}B_{r,r',s}[X_{r',s}|r\rangle\langle s|\widetilde{\rho}_{1}(t)+X_{r',r}\widetilde{\rho}_{1}(t)|r\rangle\langle s|]+\sum_{r,r',s',s}B'_{r,r',s',s}(X_{r,r'}+X_{s',s})[|r\rangle\langle r'|\widetilde{\rho}_{1}(t)|s'\rangle\langle s|], \label{master}
\end{multline}
with the initial condition $\widetilde{\rho}_{1}(0)=|n\rangle\langle n'|$.  {   We have dropped the subscript $nn'$ as the equation itself does not depend on $nn'$; only the initial condition does. } 
$|r\rangle$, $|r'\rangle$, $|s'\rangle$ and $|s\rangle$ are the
adiabatic states evaluated at $\mathbf{Q=0}.$ $B_{r,r',s}=\sum_{j}A_{r,r',\mathbf{0}}^{j}A_{r',s,\mathbf{0}}^{j}$;
$B'_{r,r',s',s}=\sum_{j}A_{r,r',\mathbf{0}}^{j}A_{s',s,\mathbf{0}}^{j}$
and $\omega_{rr'}=(\varepsilon_{r}^0-\varepsilon_{r'}^0)/\hbar$.
\begin{equation}
X_{r,r'}=\begin{cases}
\frac{\pi}{\hbar}\frac{J(\omega_{rr'})\omega_{rr'}^{2}}{e^{\beta\hbar\omega_{rr'}}-1} & \mbox{if }\omega_{rr'}>0\\
\frac{\pi}{\hbar}\frac{J(\omega_{r'r})\omega_{r'r}^{2}}{1-e^{-\beta\hbar\omega_{r'r}}} & \mbox{if }\omega_{r'r}>0.
\end{cases}
\end{equation}
Here $J_{j}(\omega)$ is the spectral density defined as
\begin{equation}
J_{j}(\omega)=\sum_{k}\frac{m_{jk}\nu_{jk}^{2}}{2\omega_{jk}}\delta(\omega-\omega_{jk}).
\end{equation}
Note that in the following we will assume that all $J_{j}(\omega)$
are the same, and denote it by $J(\omega)$, though our analysis is
valid for arbitrary $J_{j}(\omega)$.   It is worth mentioning that the method that we have used to derive the master equation is quite well known and has been used in many papers (see for example \cite{KuehnSundstroem1997,Renger&May1998}).  For deriving the master equation we use their methods.  However, our procedure differs from these papers in one crucial aspect.   For example, in \cite{KuehnSundstroem1997} the Hamiltonian is (following the notation \cite{KuehnSundstroem1997}),
\begin{equation}
H_{tot}(t) = H_{ex} + \delta H_{ex}+ H_{env} + H_{f}(t),\label{Kuehnhamiltonian}
\end{equation}
and the states $|\alpha\rangle$ that are used to derive the master equation are eigenfunctions of the
Hamiltonian $H_{ex}$ and have no dependence on the environment. They are taken to be delocalized states
of the excitonic system and all calculations are done using these. In comparison,
in our calculations, the states $|m(\textbf{Q})\rangle$ that we use are dependent on the co-ordinates of
the environment and are adiabatic eigenfunctions of the electronic Hamiltonian $H_{el}+H_{el-ph}(\mathbf{Q})$.  They have a parametric dependence on $\textbf{Q}$.
This $\textbf{Q}$ dependence makes them difficult to work with, and hence we had to introduce the
operator $T(\mathbf{Q})$.   Introducing this enabled us to  split the Hamiltonian as in Eq. (\ref{eq:Happroximate-1}) and then analyze the time evolution easily.  In the notations of reference \cite{KuehnSundstroem1997}, this means that we are working with eigenstates of $H_{ex}+\delta H_{ex}$ and not with eigenfunctions of $H_{ex}$.

\section{Application to the FMO Complex}

We now apply the above formalism to a monomer in the trimeric FMO complex with sites denoted as $1,2,3,...,7.$
We use the Hamiltonian used by Nalbach \textit{et al.} \cite{Nalbach:2011},
who have recently performed exact numerical calculations using this
Hamiltonian. The electronic part of their Hamiltonian is \begin{widetext}
\begin{equation}
\frac{H_{el}}{cm^{-1}}=\left[\begin{array}{ccccccc}
240 & -87.7 & 5.5 & -5.9 & 6.7 & -13.7 & -9.9\\
 & 315 & 30.8 & 8.2 & 0.7 & 11.8 & 4.3\\
 &  & 0 & -53.5 & -2.2 & -9.6 & 6.0\\
 &  &  & 130 & -70.7 & -17.0 & -63.3\\
 &  &  &  & 285 & 81.1 & -1.3\\
 &  &  &  &  & 435 & 39.7\\
 &  &  &  &  &  & 245
\end{array}\right].\label{matrix}
\end{equation}
\end{widetext} We have written only the part of the matrix along
the diagonal and above it. The eigenstates of the Hamiltonian are
denoted as $|m>$ with $m=1,2,\ldots7$ with respective
energies $\varepsilon_m^0$. Approximating $\varepsilon_m (\mathbf{Q}) \approx \varepsilon_m^0+\left[\nabla_{\mathbf{Q}}\varepsilon_{m}(\mathbf{Q})\right]_{\mathbf{Q=0}}.\mathbf{Q}$ ,
we get \begin{widetext}
\begin{multline}
\rho_{bc}(t)=\sum_{m,m',n\;\epsilon\;\{\psi_{i}^{0}\}}\langle b|m\rangle\langle n|a\rangle\langle a|n\rangle\langle m'|c\rangle p_{mm',nn}(t)\\
+\sum_{m,m',n',n\;\epsilon\;\{\psi_{i}^{0}\}}\langle b|m\rangle\langle n|a\rangle\langle a|n'\rangle\langle m'|c\rangle e^{-i\varepsilon_{nn'}t/\hbar}e^{-\phi_{   nn'}(t)}p_{mm',nn'}(t), \label{ImpEquation}
\end{multline}
where $\varepsilon_{nn'}=\varepsilon_{n}^0-\varepsilon_{n'}^0$, $\phi_{n,n'} (t)= Re( \phi_{n,n'} (t)+i Im(\phi_{n,n'} (t))$ and
\begin{equation}
Re(\phi_{n,n'}(t))=\frac{1}{\hbar}\int_{0}^{\infty}d\omega J(\omega)\frac{1-\cos(\omega t)}{\omega^{2}}\coth\left(\frac{\beta\hbar\omega}{2}\right)\sum_{j=1,2,...,7}\left(\frac{\partial\varepsilon_{n}}{\partial Q_{j}}-\frac{\partial\varepsilon_{n'}}{\partial Q_{j}}\right)_{Q_{j}=0}^{2},\label{eq:Rephi}
\end{equation}
and
\begin{equation}
Im(\phi_{n,n'}(t))=\frac{1}{\hbar}\int_{0}^{\infty}d\omega J(\omega)\frac{\sin(\omega t)-\omega t}{\omega^{2}}\sum_{j=1,2,...,7}\left(\left(\frac{\partial\varepsilon_{n}}{\partial Q_{j}}\right)_{Q_{j}=0}^{2}-\left(\frac{\partial\varepsilon_{n'}}{\partial Q_{j}}\right)_{Q_{j}=0}^{2}\right).\label{eq:Imphi}
\end{equation}
{   Equations (\ref{master}) and (\ref{ImpEquation}) and  together with (\ref{eq:Rephi}) and (\ref{eq:Imphi}) form the basic equations of our calculation.  If one puts $\phi_{nn'}$ in Eq. (\ref{ImpEquation}) equal to zero, and imagined $\overline{H}_{na}$ to be the only perturbation, then the our method of calculation would be same as that of \cite{KuehnSundstroem1997,Renger&May1998} .}
\end{widetext}

First we will consider the Drude
spectral density as it has been extensively used for the theoretical
modeling of excitation energy transfer in the FMO complex. It is given by
\begin{equation}
J(\omega)=J_{D}(\omega)=\frac{2\lambda}{\pi}\frac{\omega\omega_{c}}{\omega^{2}+\omega_{c}^{2}},
\end{equation}
where $\lambda$ is the reorganization energy and $\alpha=\frac{\lambda}{\hbar\omega_{c}}$. In our calculations, we consider $\lambda=35\ cm^{-1}$
and $\omega_{c}^{-1}=50\ fs.$ Expressions for the integrals in Eq.
(\ref{eq:Rephi}) and Eq. (\ref{eq:Imphi}) are given in the Appendix \ref{AppendixA}. We also performed calculations for the more realistic spectral density determined
by Adolphs and Renger for the FMO complex \cite{AdolphsRenger:2006}. Their spectral density
has a discrete mode too and is given by
\begin{equation}
J(\omega)=J_{AR}(\omega)=\omega^{2}S_{0}g_{0}\left(\omega\right)+J_{dm}(\omega),
\end{equation}
where $J_{dm}(\omega)=\omega^{2}S_{H}\delta\left(\omega-\omega_{H}\right)$ and
\[
g_{0}\left(\omega\right)=6.105\times10^{-5}\times\frac{\omega^{3}}{\omega_{1}^{4}}e^{-\sqrt{\frac{\omega}{\omega_{1}}}}+3.8156\times10^{-5}\times\frac{\omega^{3}}{\omega_{2}^{4}}e^{-\sqrt{\frac{\omega}{\omega_{2}}}},
\]
with $S_{0}=0.5$, $S_{H}=0.22$, $\omega_{H}=180\ cm^{-1}$, $\omega_{1}=0.575\ cm^{-1}$
and $\omega_{2}=2\ cm^{-1}$. Nalbach and coworkers \cite{Nalbach:2011}, use a spectral density in which the discrete mode has a broadening of $\gamma_p$, so that
\begin{equation}
J_{dm}(\omega)=\omega_{H}^{2}S_{H}\frac{1}{\pi}\frac{\gamma_{p}}{\left(\omega-\omega_{H}\right)^{2}+\gamma_{p}^{2}}. \label{JAR}
\end{equation}
We use a slightly different way of broadening the delta function, as this makes the calculations of the decoherence expressions analytical.
We take the contribution to the spectral density
from the discrete mode to be
\begin{equation}
J_{dm}(\omega)=\omega\omega_{H}S_{H}\frac{1}{\pi}\frac{\gamma_{p}}{\left(\omega-\omega_{H}\right)^{2}+\gamma_{p}^{2}}.
\end{equation}
We have performed calculations
for  $\gamma_{p}=\ 1\ cm^{-1}$.   There is negligible difference between the two densities as may be confirmed by making plots of them.
To evaluate the matrix elements \begin{equation}\widetilde{\rho}_{1,mm'}(t)=\langle m|\rho_1(t)|m'\rangle, \end{equation}  we use the following equations
respectively for populations and coherences, which are obtained from Eq. (\ref{master}), by calculating  matrix elements of this equation.  

\begin{widetext}
\begin{equation}
\dot{\widetilde{\rho}}_{1,mm}(t)=-\left(\sum_{n,n\neq m}\Gamma_{nm}\right)\widetilde{\rho}_{1,mm}(t)+\sum_{n,n\neq m}\Gamma_{mn}\widetilde{\rho}_{1,nn}(t)\label{eq:population}
\end{equation}

\begin{equation}
\dot{\widetilde{\rho}}_{1,mn}(t)=-\frac{1}{2}\left(\Gamma_{mn}+\Gamma_{nm}\right)\widetilde{\rho}_{1,mn}(t),\label{eq:coherence} \hspace{0.5cm} n\neq m.
\end{equation}
\end{widetext} Therefore, we would have a matrix equation of the
following form:
\[
\widetilde{\rho}_1\left(t\right)=e^{\mathbf{\Gamma}t}\widetilde{\rho}_1\left(0\right),
\]
where $\mathbf{\Gamma}$ is a $7\times7$ matrix containing the rate
constant elements $\Gamma_{mn}$. $\widetilde{\rho}_1\left(t\right)$
is the $N\times N$ matrix of populations and coherences at any time
$t$.

We have $\omega_{mn}=\frac{\varepsilon_{m}^0-\varepsilon_{n}^0}{\hbar}=\omega_{m}-\omega_{n}$.
If $\varepsilon_{m}^0>\varepsilon_{n}^0$ , this gives $\Gamma_{nm}=e^{\beta\hbar\omega_{mn}}\Gamma_{mn}$
with
\begin{eqnarray}
\Gamma_{mn}=\frac{2\pi}{\hbar}\frac{J(\omega_{mn})}{e^{\beta\hbar\omega_{mn}}-1}\sum_{j}\left(\langle n|j\rangle\langle j|m\rangle\right)^{2}.
\end{eqnarray}
  It is worth stressing that the eigenfunctions $|m\rangle$ and $|n\rangle$ are just the delocalized eigenstates of the system  at $\mathbf{Q}=\mathbf{0}$ and hence our approach in calculation of the relaxation rate resembles the approaches of \cite{KuehnSundstroem1997,Renger&May1998}. However, our procedure differs from previous approaches \cite{KuehnSundstroem1997,Renger&May1998} in two respects:  (1) The use of adiabatic eigenstates has enabled us to split the effects of interaction with the environment into two parts.  (2) The term responsible for the major part of decoherence is accounted for separately in a non-perturbative fashion and the non adiabatic term $\overline{H}_{na,\mathbf{0}}$ which causes population relaxation is accounted for using the methods of \cite{KuehnSundstroem1997,Renger&May1998}. Further, it is important to note that the perturbation that is used in the calculation of the master equation is not $\delta H_{ex}$ (in the notation of \cite{KuehnSundstroem1997}, which is reproduced in our Eq. (\ref{Kuehnhamiltonian})), but $\overline{H}_{na,\mathbf{0}}$ that is given in Eq. (\ref{eq:HforMasterEquation}).  
\section{Results and Discussions}

\begin{figure}[!tph]
\includegraphics[width=0.8\linewidth]{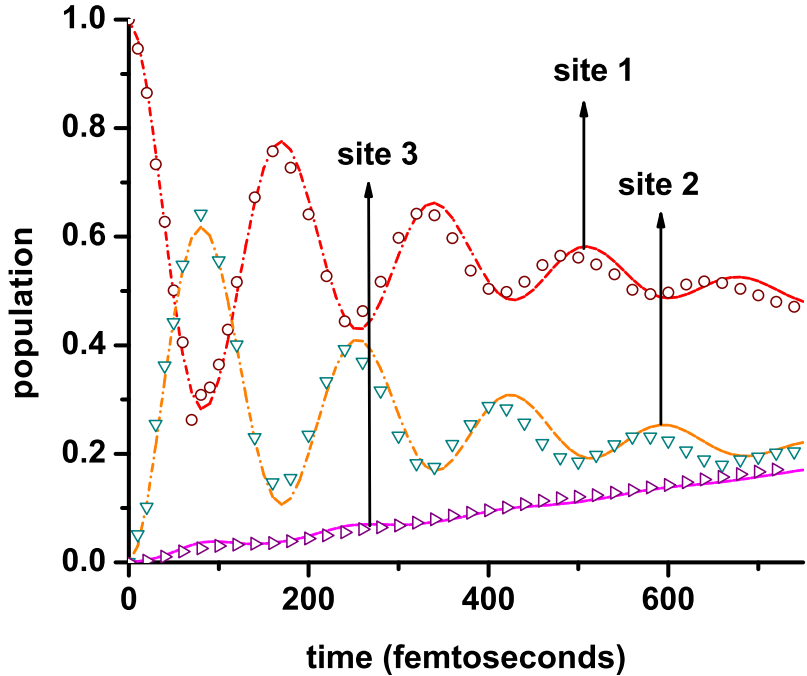}
\caption{Calculations for the spectral density  $J_{D}(\omega)$. Initial excitation is on site 1 and temperature is 77 K. Comparison of our results (dashed
lines) with those of Nalbach \textit{et al.}\cite{Nalbach:2011} (denoted by symbols) for chromophores
1,2 and 3.}

\label{Fig1}
\end{figure}

We discuss below the results for the seven  level FMO
complex using the analytical expressions obtained above. We
consider the initial excitation to reside at either site 1 or site
6 as they are located nearest to the chlorosome antenna. We provide
an instance of how well our method compares with the exact numerical
calculations by Nalbach \textit{et al.}
\cite{Nalbach:2011}. In Fig. \ref{Fig1} we give results for the
case where the initial excitation is at site 1. The figure shows the
probability of finding the excitation at sites 1, 2 and 3, at 77 K.
The agreement of our method with the exact results is seen to be very
good. We have compared for the other cases when the initial excitation
is at site 6 and at two different temperatures of 77 K and 300 K
(figures are not given, to save space). The agreement is excellent
for all the cases.

\begin{figure}[!tph]
\includegraphics[width=0.8\linewidth]{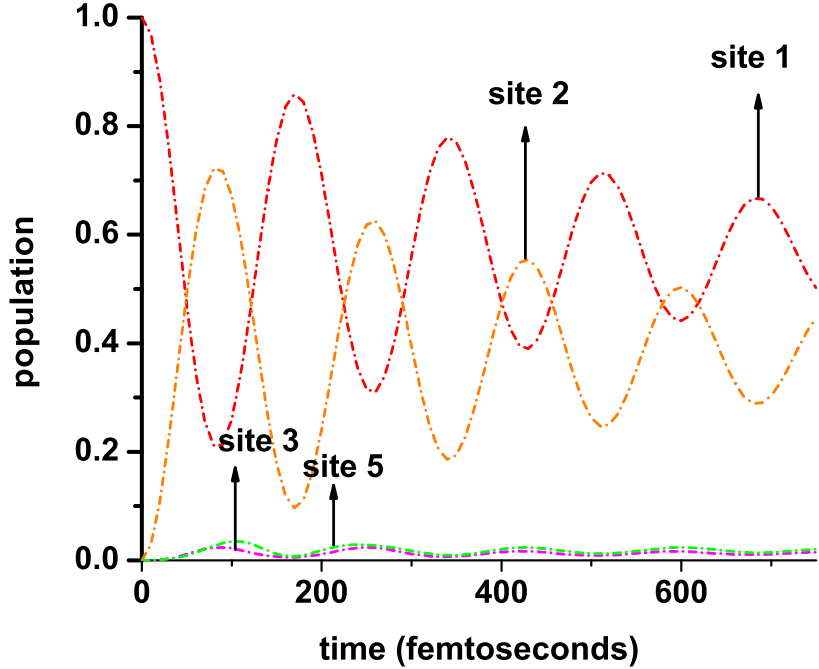}\caption{Results for decoherence caused by the environment when the spectral density is   $J_{D}(\omega)$ and $T=77K$. Initial excitation is put on  site 1. The curves for sites 4, 6 and 7 are not shown as the population at these sites are less than that of site 3 at all times.}

\label{Fig2}
\end{figure}

\begin{figure}[!tph]
\includegraphics[width=0.8\linewidth]{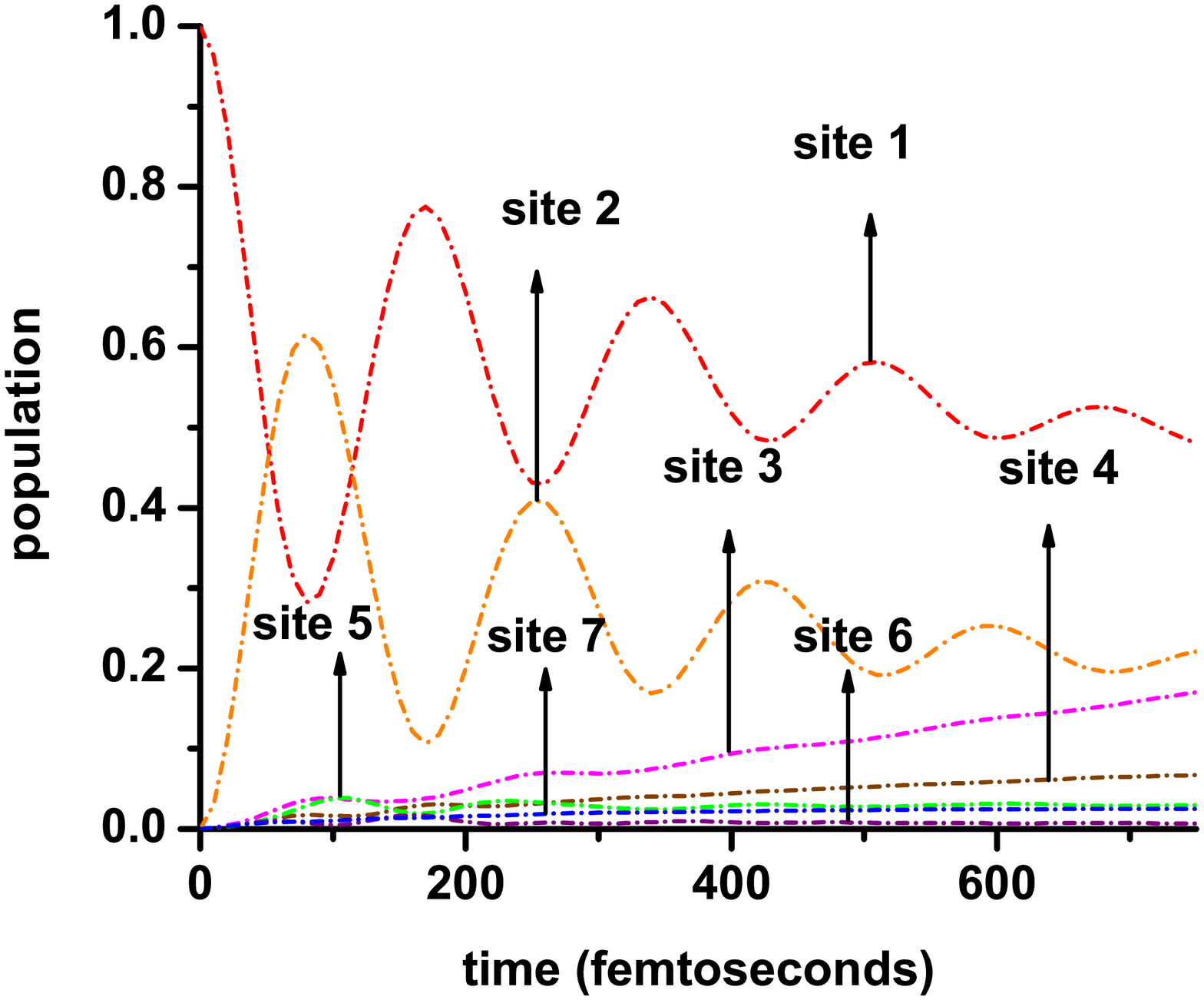}\caption{Calculations for spectral density $J_{D}(\omega)$. Initial excitation is taken to be on site 1  and temperature to be  77 K.  Curves  shown include effects of  decoherence and population relaxation
due to the environment.}
\label{Fig3}
\end{figure}

With our approach, it is possible to look at the
effect of decoherence and population relaxation separately. First,
we consider the case when the initial excitation is put on site 1
and the environment is absent. The excitation mostly oscillates between
sites 1 and 2 and a very small amount of the total excitation is transferred
to the other sites, including site 3 which has the least energy. This
is due to the strong off-diagonal coupling in the Hamiltonian between
chromophores 1 and 2. If we confine the initial excitation to chromophore
6 and again consider the environment to be absent, the excitation
oscillates among the sites 6, 5, 4 and 7 but there is again no appreciable
transfer to chromophore 3 which is closest to the reaction centre.
We now investigate the impact of decoherence at two temperatures,
77 K and 300 K, respectively. Figure \ref{Fig2} has the initial excitation
on 1 at 77 K and shows only the decoherence effects induced by the
environment. We have coherent oscillations of the excitation mostly
between chromophores 1 and 2 with no significant transfer elsewhere.
The oscillatory behaviour of the excitation gets somewhat damped,
due to decoherence, at timescales of $\backsim 750\ fs$. However,
once we include the environment - induced population relaxation effects
along with decoherence (Figure \ref{Fig3}), the oscillations at sites
1 and 2 get damped rapidly (the oscillations reduce significantly
at $\sim750\ fs$) and their amplitudes decrease significantly. Chromophore
3 now has a significant amount of excitation transferred to it and
it exhibits oscillations only upto $300\ fs$. As there was no appreciable
transfer to site 3 in the absence of population relaxation, this is
just the environment assisted transport, suggested previously by other
authors. There is also some population transfer to chromphore 4 which
hardly shows any oscillatory behaviour.

\begin{figure}[!tph]
\includegraphics[width=0.8\linewidth]{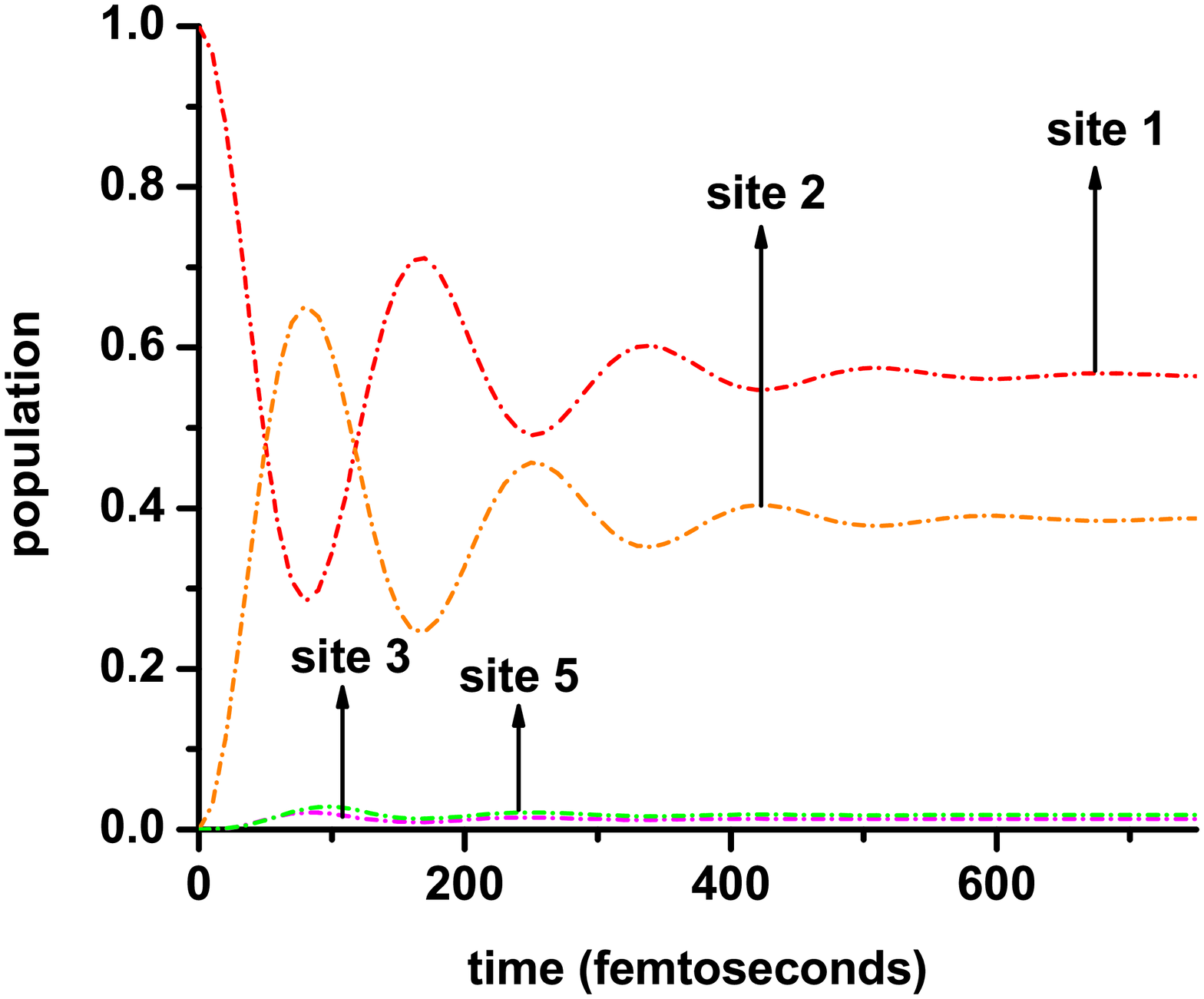}
\caption{Calculations at 300 K,  for the spectral density  $J_{D}(\omega)$. Initial excitation is assumed to be on site 1.  Calculations account only for decoherence and does not include population relaxation. The curves for sites 4, 6 and 7 are not shown as the population at these sites are less than that of site 3 at all times.}

\label{Fig4}
\end{figure}

\begin{figure}[!tph]
\includegraphics[width=0.8\linewidth]{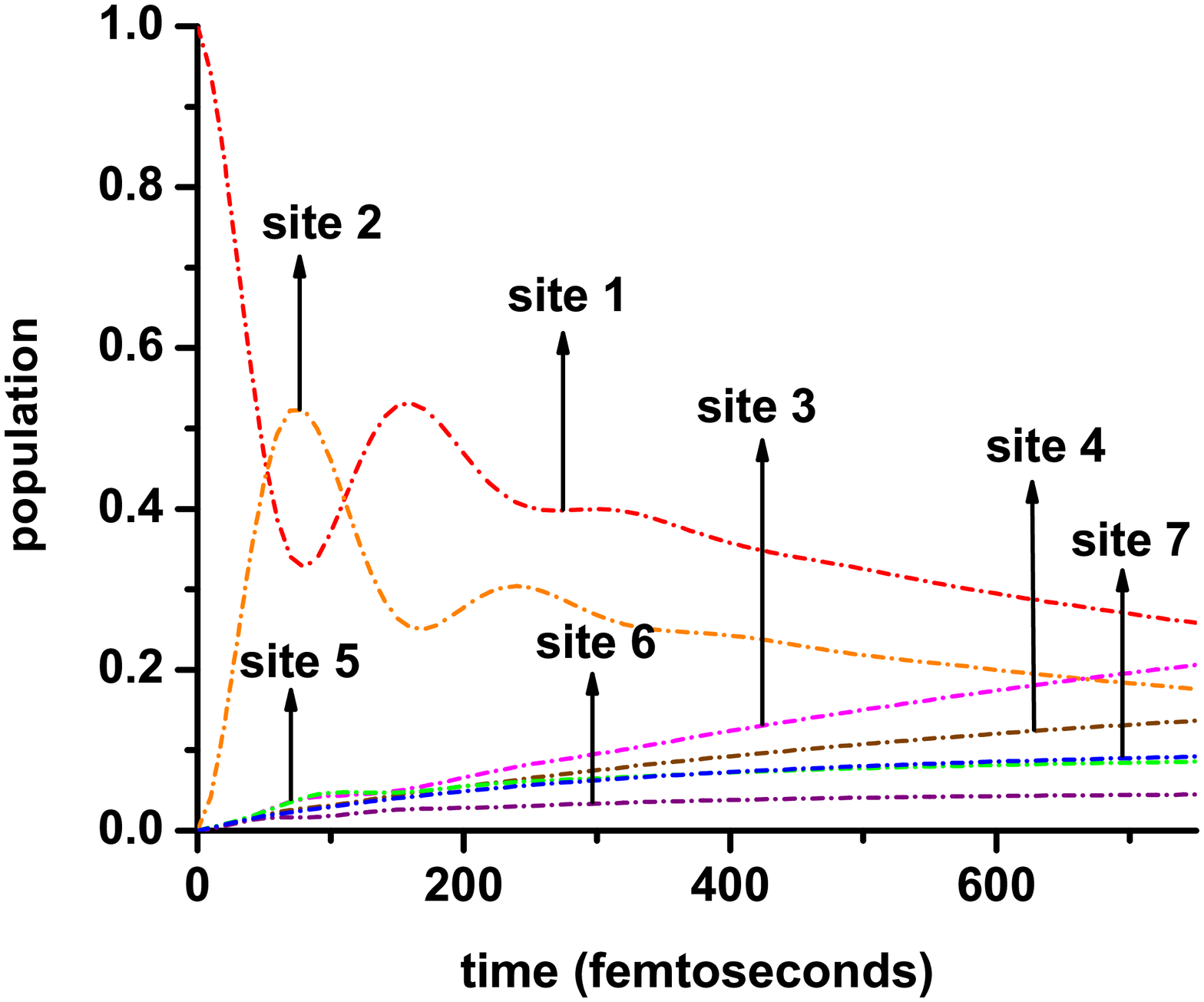}
\caption{Calculations at 300 K, for spectral density  $J_{D}(\omega)$. Initial excitation is assumed to be on site 1. Results include both decoherence and population
relaxation due to the environment.}

\label{Fig5}
\end{figure}

We now consider the environment - induced effects
at a higher temperature, 300 K. Figure \ref{Fig4} considers only
the decoherence effects on population evolution at the different chromophores
with the initial population at site 1. A higher temperature causes
severe damping in the oscillatory behaviour of the excitation, which
is again mostly confined to sites 1 and 2. The other sites do not
receive any appreciable amount of energy. But on inclusion of the
population relaxation effects as well (Figure \ref{Fig5}), we see
the excitation getting transferred to other sites with significant
transfers to sites 3 and 4. However the coherent oscillations get
damped and disappear around $350\ fs.$

\begin{figure}[!tph]
\includegraphics[width=0.8\linewidth]{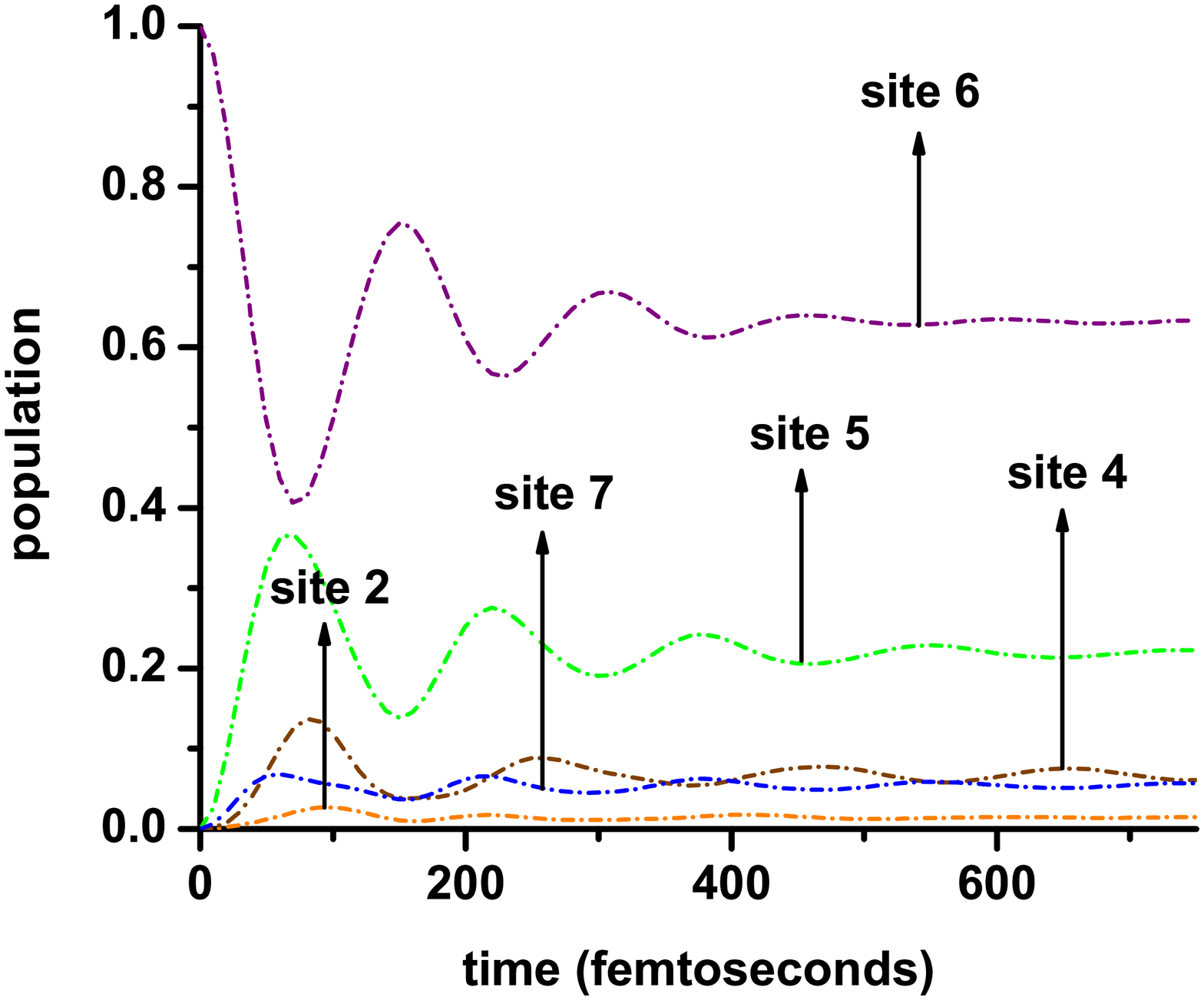}
\caption{Calculations for the spectral density $J_{D}(\omega)$ and temperature 77 K. Initial excitation is assumed to be on site 6. The curves include only the effects of decoherence and do not include effects of population relaxation. The curves for sites 1 and 3 are not shown as the population at these sites are less than that of site 2 at all times.}
\label{Fig6}
\end{figure}

\begin{figure}[!tph]
\includegraphics[width=0.8\linewidth]{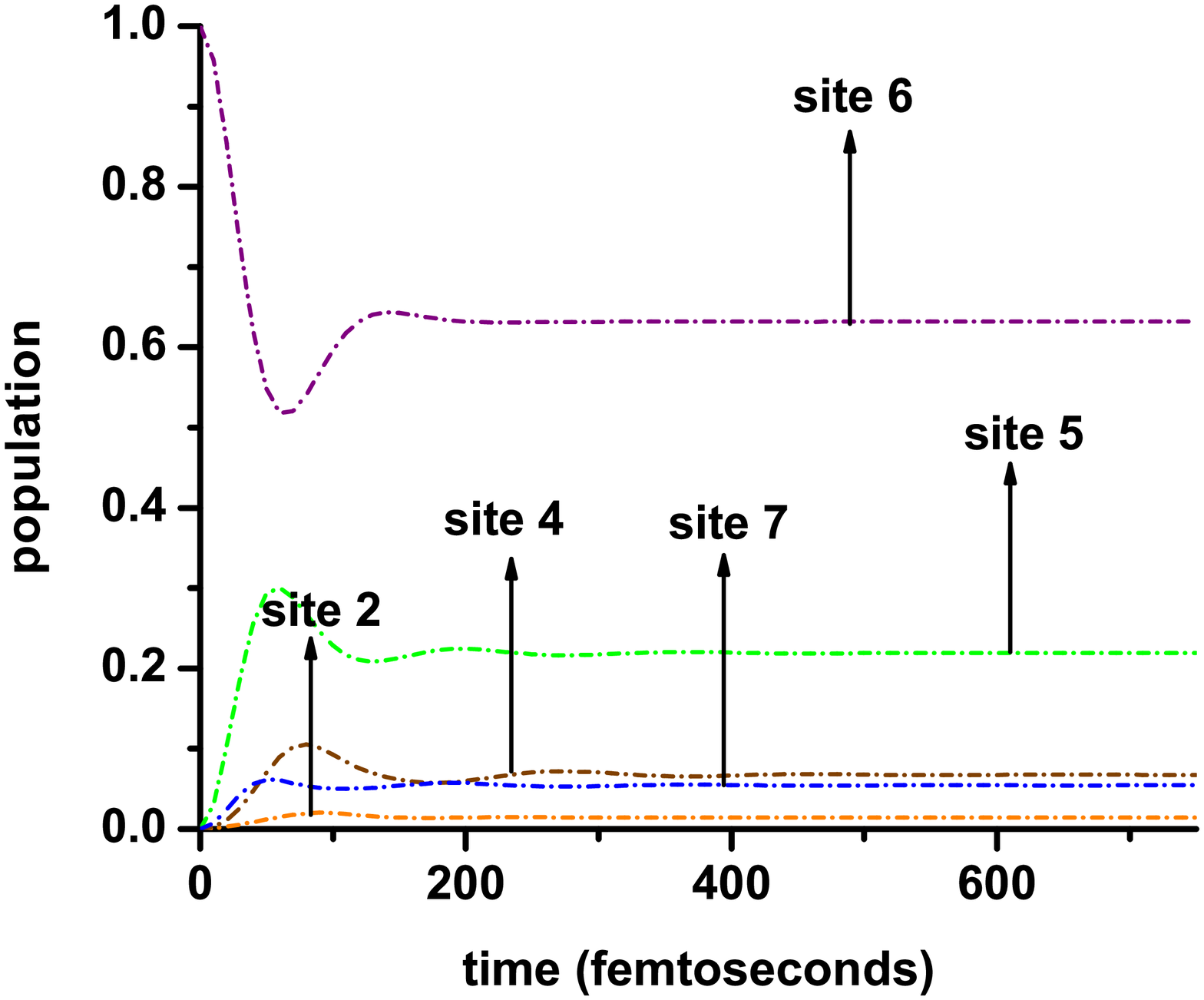}
\caption{Calculations for the spectral density $J_{D}(\omega)$ and temperature 77 K. Initial excitation is assumed to be on site 6. The curves include only the effects of decoherence and do not include effects of population relaxation. The curves for sites 1 and 3 are not shown as the population at these sites are less than that of site 2 at all times.}

\label{Fig7}
\end{figure}

We now consider the initial excitation to be at site 6 and investigate the decoherence effects due to the environment,
with and without population relaxation, at the two temperatures. Figures
\ref{Fig6} and \ref{Fig7} show only the decoherence effects at 77
K and 300 K respectively. In both the cases, the excitation mostly
oscillates between chromophores 6, 5, 4 and 7 with no significant transfer
elsewhere, as in the case where the environment was considered absent.
However, higher the temperature we consider, greater would be the
damping and the coherent oscillations would disappear faster. At 77
K, the oscillations, though damped, persist upto $\backsim500\ fs$.
However, at 300 K, the oscillations disappear around $\backsim200\ fs$.

\begin{figure}[!tph]
\includegraphics[width=0.8\linewidth]{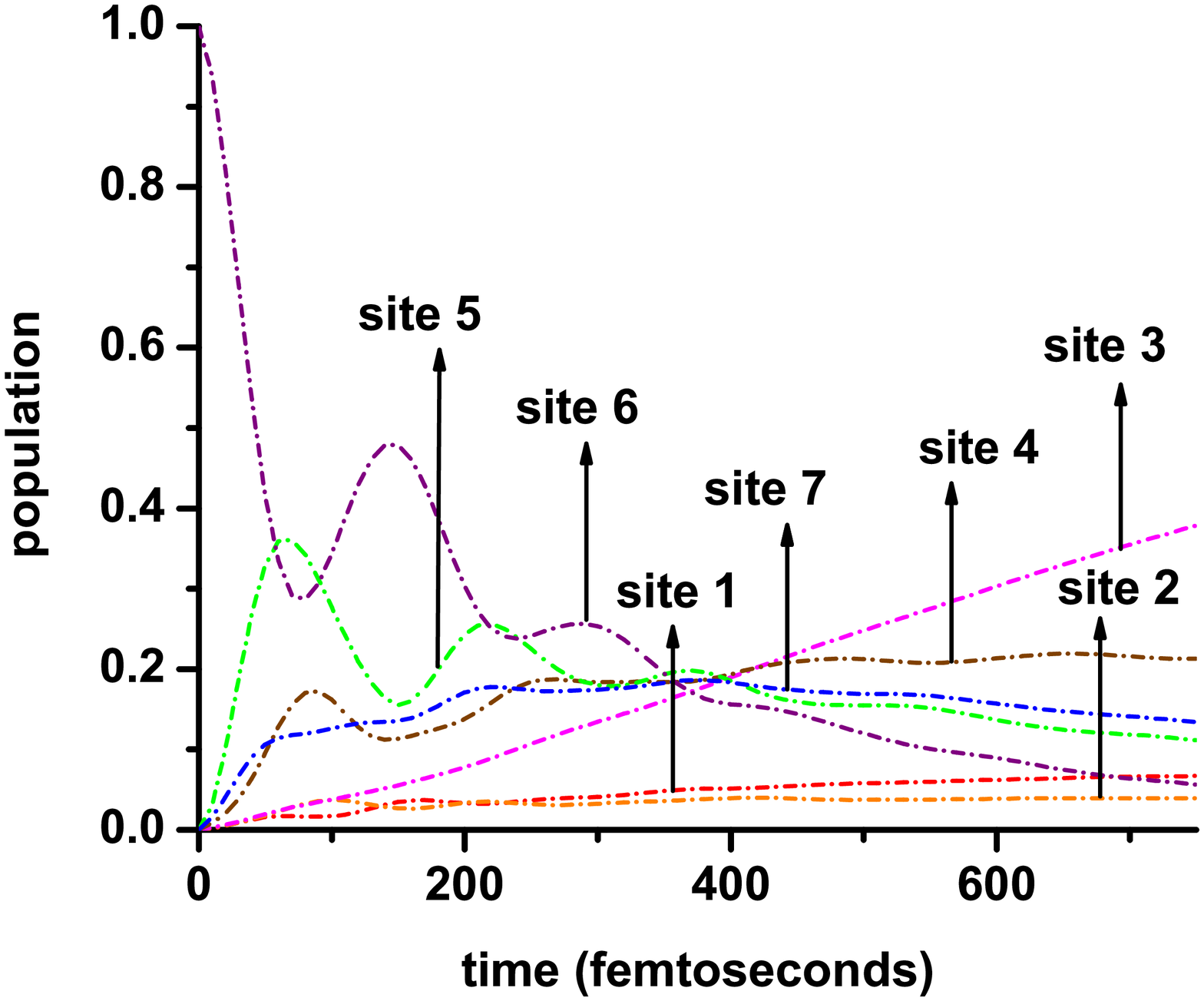}
\caption{Calculations for the spectral density $J_{D}(\omega)$ and temperature 77 K. Initial excitation is assumed to be on site 6.
The curves include effects of  both decoherence and population relaxation
due to the environment.}

\label{Fig8}
\end{figure}

\begin{figure}[!tph]
\includegraphics[width=0.8\linewidth]{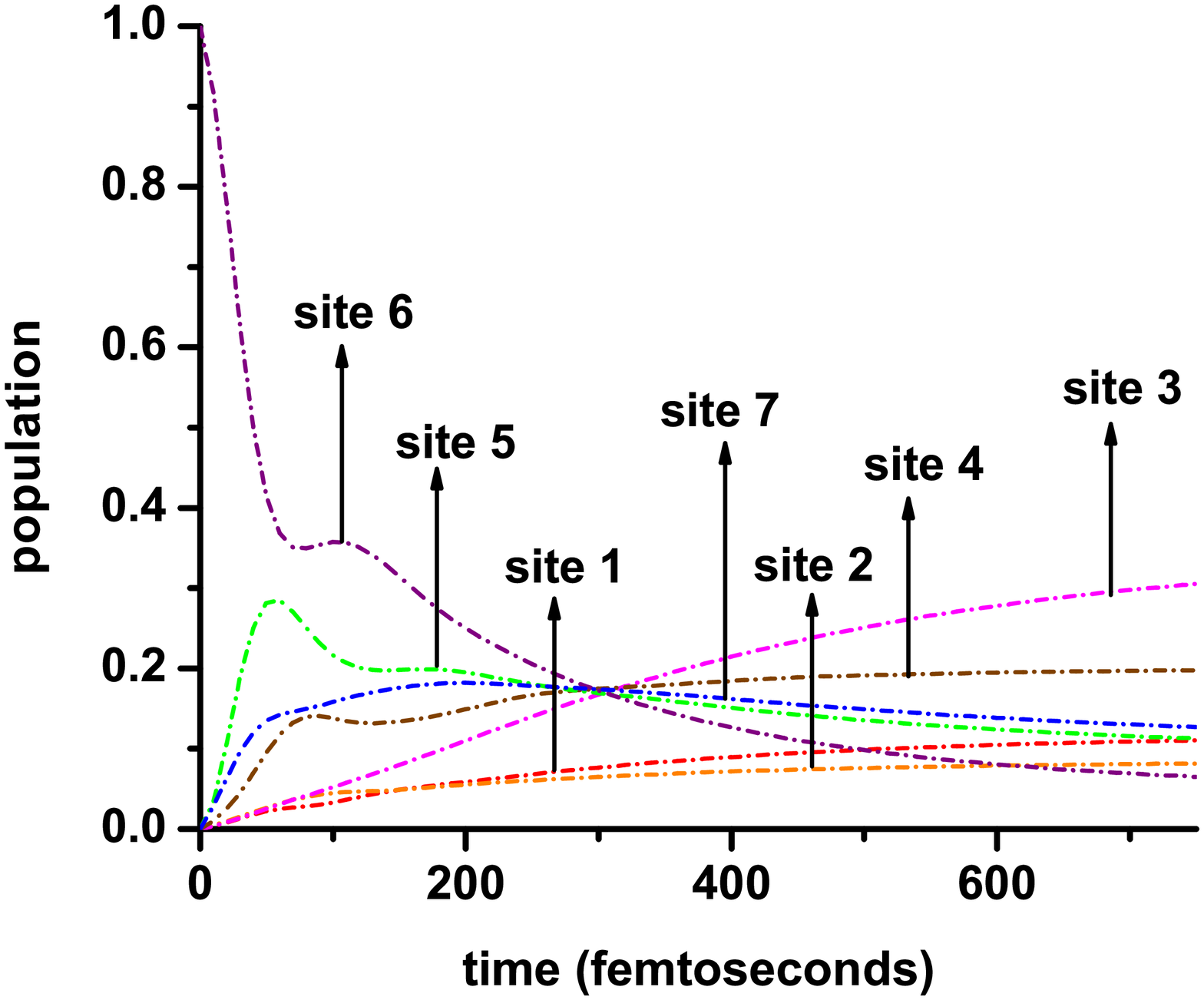}
\caption{Calculations for the spectral density $J_{D}(\omega)$ and temperature 300 K. Initial excitation is assumed to be on site 6.
The curves include effects of  both decoherence and population relaxation
due to the environment.}

\label{Fig9}
\end{figure}

Inclusion of the population relaxation effects (Figures
\ref{Fig8} and \ref{Fig9}) results in faster damping and more importantly,
in transfer of the population to the other sites, most importantly
to chromophore 3, which is located closest to the reaction centre
and chromophore 4. Again, thermal equilibrium is attained faster at
higher temperatures.

\begin{figure}[!tph]
\includegraphics[width=0.8\linewidth]{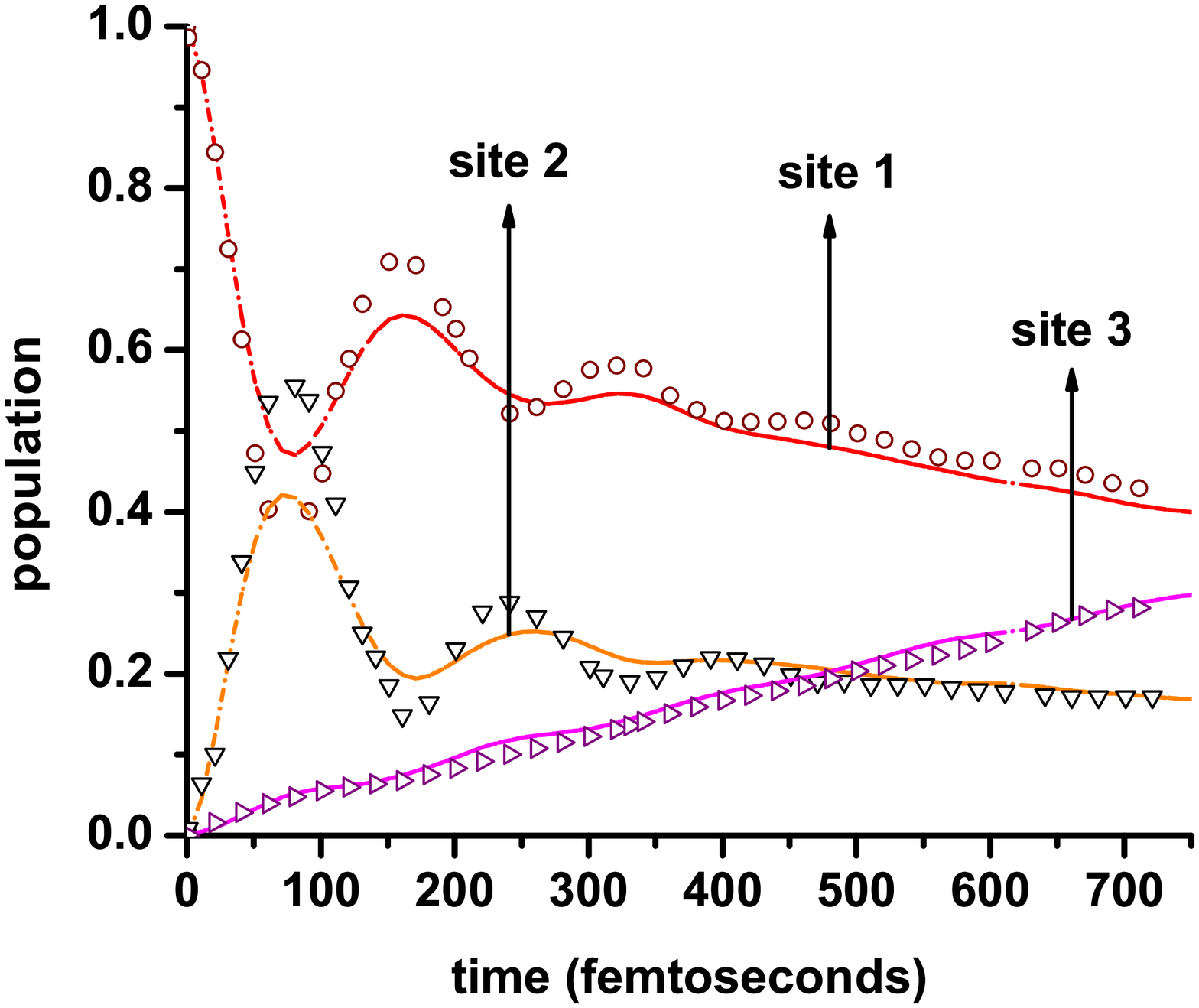}
\caption{Calculations for the spectral density  $J_{AR}(\omega)$.  We have taken  $\gamma_{p}=1\ cm^{-1}$ and temperature to be 77 K. Initial excitation is taken to be on site 1.    The curves include effects of decoherence and population relaxation
due to the  environment. The figure compares our results (lines) with those
of Nalbach \textit{et al.}\cite{Nalbach:2011} (denoted by symbols) for chromophores 1, 2 and
3 and  the agreement is found to be good.}
\label{Fig10}
\end{figure}

We now present the calculations for the more realistic spectral
density $J_{AR}(\omega)$ suggested by Adolphs and Rengers \cite{AdolphsRenger:2006}, for which exact numerical
calculations have been done  by Nalbach and coworkers \cite{Nalbach:2011}. We show only the results which arise from
decoherence and population relaxation considered together. Again, we provide a comparison of our results with them to show how
well the method works, despite the approximations in Figure \ref{Fig10}.
 The agreement is seen to be very good.

\begin{figure}[!tph]
\includegraphics[width=0.8\linewidth]{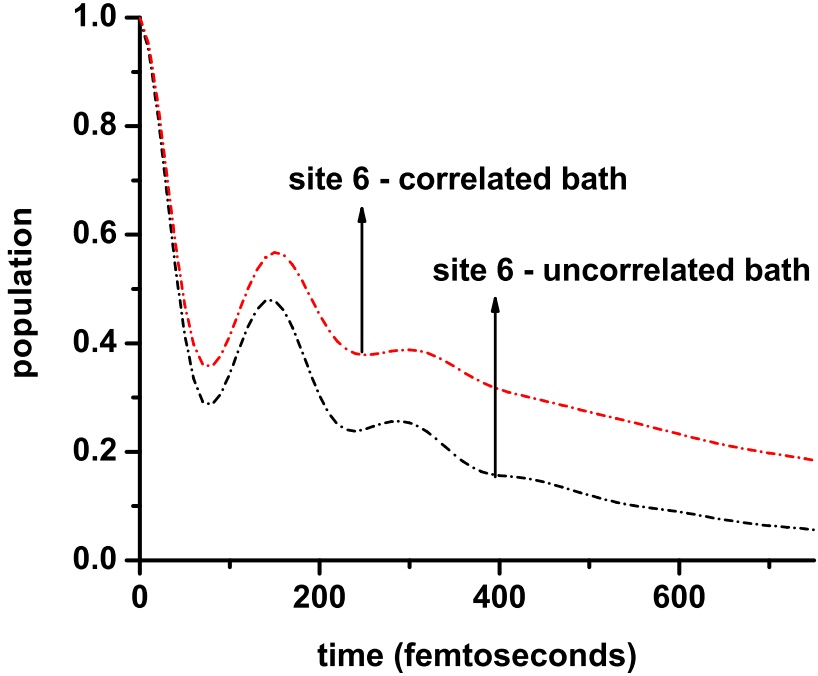}
\caption{Calculations for the spectral density  $J_{D}(\omega)$.  Initial excitation assumed to be on site 6 and the temperature is taken to be 77 K.   Results are given for the population on site 6, for both correlated and uncorrelated baths. }

\label{Fig11}
\end{figure}

\begin{figure}[!tph]
\includegraphics[width=0.8\linewidth]{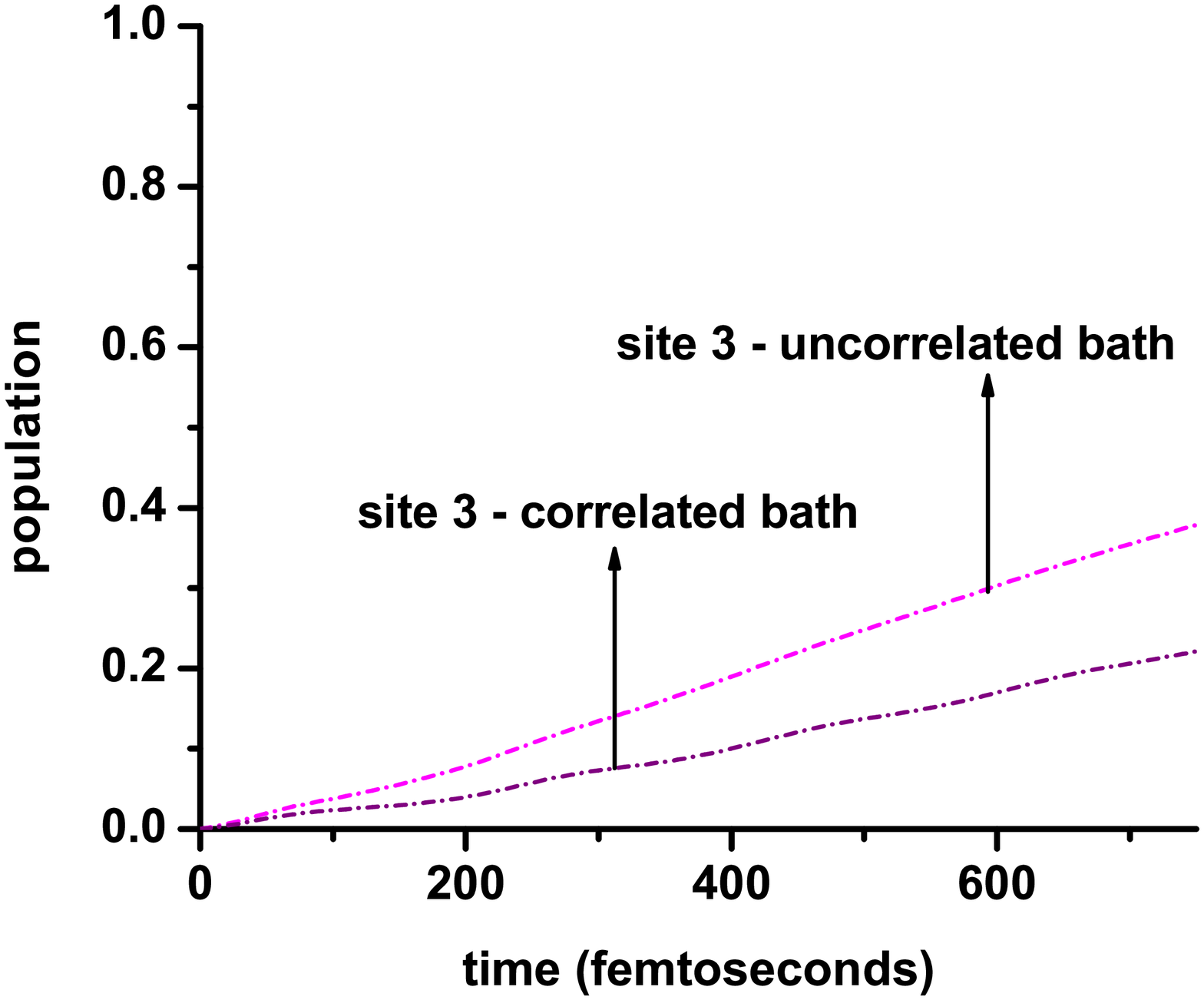}
\caption{Calculations for the spectral density  $J_{D}(\omega)$.  Initial excitation is assumed to be on site 6 at 77 K.   The plots show probability of transfer to site 3, for correlated and uncorrelated baths. }

\label{Fig12}
\end{figure}

\section{Correlated Bath}

It is also possible to investigate,
how the presence of correlations among the bath degrees of freedom
affects the process of light-harvesting \cite{Sarovar:2011}.  A numerical investigation of the effect of correlations on the simple two level system was done by Nalbach and coworkers \cite{NalbachNJPCB2010}.  We investigate the effect of correlations for the seven level system using our method.  Reference \cite{Sarovar:2011} suggests three different models for correlations.
 The first has no correlations, which is what we have studied in detail in this paper.  The second one has  correlations between neighboring sites while the third has correlations decay exponentially with distance.    We have studied the second model, as we wish to see what happens if correlations are included.  Physically, it seems unlikely that there are correlations, as the sites are rather apart in space.   It should be pointed out that recent simulations \cite{Olbrich:2011ez} found no correlations in the fluctuations of the energies at different sites. A quantitative
measure of the spatial correlation is given by the correlation matrix
$C$ of Eq. (\ref{CMatrix}). Following reference \cite{Sarovar:2011},
we use $C_{12}=C_{21}=C_{56}=C_{65}=0.9;C_{45}=C_{54}=C_{47}=C_{74}=0.4.$
Figures \ref{Fig11} and \ref{Fig12} compare the evolution of population
for correlated and uncorrelated bath for two specific cases: a) population
evolution at site 6 when the initial excitation is at site 6 at 77
K and b) population evolution at site 3 when the initial excitation
is at site 6 at 77 K. The spectral density used here is the Drude spectral density.
From Figure \ref{Fig11}, we see that in the
absence of correlations, the population at site 6 reaches thermal
equilibrium faster. However, we would be more interested in the excitation
transfer to site 3 as it sits closest to the reaction centre. In Figure
\ref{Fig12}, the excitation is transferred much faster and hence
more efficiently to site 3 in the absence of correlations.
 It has been recently suggested that vibrational coherences may be responsible for the oscillations that are observed in the experiments with the FMO complex \cite{PlenioNature2013,TiwariPNAS2013,PolyutovCP2012,ChristenssonJPCB2012}.  This requires that one or more vibrations exchange energy back and forth with electronic excitations. Proper accounting for this requires that these vibrations should be accounted for in a more detailed manner than we have done.  We are currently investigating the extension of our approach to account for this.  Also, it has been suggested that the initial state that is produced in the experiment may not be on a single site, but it may be delocalized over two (or more) sites.   In our approach this is quite easy to account for - all that one needs is to express the initial state in terms of the delocalized eigenfunctions.   We are currently investigating such states using the method.

\section{Conclusions}

In this paper, we have introduced an analytical technique
for treating coherent wavelike excitation transfer. We wish to emphasize
that unlike the existing numerical techniques, the method is computationally
inexpensive and could be applied to systems with large number of states.
Again, the existing approaches mostly employ perturbative techniques.
However, our treatment, which employs mapping onto the adiabatic basis,
accounts for decoherence and population relaxation independently.
Decoherence is accounted for non-perturbatively, and population relaxation
using a Markovian master equation. Results are presented for two spectral
densities: a) Drude spectral density and b) the spectral density determined
by Adolphs and Renger \cite{AdolphsRenger:2006} for the FMO complex. The results obtained suggest that if
only the environment-induced decoherence is considered, there is only
damping observed in the oscillatory nature of the excitation which
otherwise remains confined to the same sites as in the case where
we have considered the environment absent. A higher temperature would
only induce more severe damping. In other words, on confining the
initial excitation to site 1 or site 6, coherence doesn't help the
excitation propagate to site 3 which has the least energy and is located
closest to the reaction centre where the electron transfer reaction
takes place. \textit{The energy transfer to site
3 and the other sites takes place only when the population relaxation
effects are included along with coherence effects.}
Again, at higher temperatures, attainment of thermal equilibrium due
to this would be faster. Therefore, the population relaxation, which
is responsible for washing out of the coherences, is the principal
reason for the excitation to move to site 3. The coherent oscillations,
which do not persist for large enough times, remain essentially confined
to a subset of chromophores and do not help in propagating the excitation
to site 3. The method also works well when we use the spectral density suggested by Adolphs and Renger \cite{AdolphsRenger:2006}, which has a discrete mode too.  We also investigate
the effects of presence of bath correlations. The excitation transfer
to the reaction centre is more efficient in their absence. To conclude,
the extremely efficient light harvesting phenomenon is rendered so
by the dissipative influence of the environment and not its absence,
as usually suspected.

\section{Acknowledgements}

The work of K.L. Sebastian was supported by the Department of Science
and Technology, Govt. of India by the J.C. Bose fellowship program.
Pallavi Bhattacharyya thanks the Indian Institute of Science for scholarship.

\appendix
\section{Evaluating the Decoherence Terms}
\label{AppendixA}
We can evaluate $Tr_{ph}\left\{ \left(\hat{T}e^{-i\int_{0}^{t}dt_{1}\varepsilon_{n}(\mathbf{Q}(t_{1}))/\hbar}\right)\rho_{ph}(0)\left(\hat{T}^{\dagger}\, e^{i\int_{0}^{t}dt_{1}\varepsilon_{n'}(\mathbf{Q}(t_{1}))/\hbar}\right)\right\} $
in the following manner: We have from Eq. (\ref{eq:Hzerobar})
\[
\overline{H}_{0}=\sum_{m}\varepsilon_{m}(\mathbf{Q})c_{m}^{\dagger}c_{m}+\frac{1}{2}\sum_{j,k}\left\{ \frac{\widehat{p}_{jk}^{2}}{m_{jk}}+m_{jk}\omega_{jk}^{2}q_{jk}^{2}\right\}
\]
We consider upto first order in $\varepsilon_{m}(\mathbf{Q})$ with
respect to $Q$. Consequently, $\varepsilon_{m}(\mathbf{Q})=\varepsilon_{m}^0+\sum_{j}\left(\frac{\partial\varepsilon_{m}}{\partial Q_{j}}\right)_{Q_{j}=0}Q_{j}$
. Therefore, $\overline{H}_{0}=H_{0}+V\,$ where $H_{0}=\sum_{m}\varepsilon_{m}^0c_{m}^{\dagger}c_{m}+\frac{1}{2}\sum_{j,k}\left\{ \frac{\widehat{p}_{jk}^{2}}{m_{jk}}+m_{jk}\omega_{jk}^{2}q_{jk}^{2}\right\} $
and $V=\sum_{j,m}\left(\frac{\partial\varepsilon_{m}}{\partial Q_{j}}\right)_{Q_{j}=0}Q_{j}\, c_{m}^{\dagger}c_{m}$.
Now we write the following matrix element in the interaction picture
as follows:

\[
Tr_{ph}\left\{ \left(\hat{T}e^{-i\int_{0}^{t}dt_{1}\varepsilon_{n}(\mathbf{Q}(t_{1}))dt_{1}/\hbar}\right)\rho_{ph}(0)\left(\hat{T}^{\dagger}\, e^{i\int_{0}^{t}dt_{1}\varepsilon_{n'}(\mathbf{Q}(t_{1}))dt_{1}/\hbar}\right)\right\}
\]
\[
=Tr_{ph}\left\{ \left\langle n\left|e^{-i\overline{H}_{0}t/\hbar}\right|n\right\rangle \rho_{ph}(0)\left\langle n'\left|e^{i\overline{H}_{0}t/\hbar}\right|n'\right\rangle \right\}
\]
\[
=Tr_{ph}\left\{ \left\langle n\left|e^{-iH_{0}t/\hbar}e^{iH_{0}t/\hbar}e^{-i\overline{H}_{0}t/\hbar}\right|n\right\rangle \rho_{ph}(0)\left\langle n'\left|e^{i\overline{H}_{0}t/\hbar}e^{-iH_{0}t/\hbar}e^{iH_{0}t/\hbar}\right|n'\right\rangle \right\}
\]
\[
=e^{-i\varepsilon_{nn'}t/\hbar}Tr_{ph}\left\{ \left\langle n\left|e^{iH_{0}t/\hbar}e^{-i\overline{H}_{0}t/\hbar}\right|n\right\rangle \rho_{ph}(0)\left\langle n'\left|e^{i\overline{H}_{0}t/\hbar}e^{-iH_{0}t/\hbar}\right|n'\right\rangle \right\}
\]
\[
=e^{-i\varepsilon_{nn'}t/\hbar}Tr_{ph}\left\{ \left(\hat{Te^{-i\int_{0}^{t}dt_{1}V_{I}\left(t_{1}\right)/\hbar}}\,\right)\rho_{ph}(0)\left(\hat{T}^{\dagger}e^{i\int_{0}^{t}dt_{1}V_{I}\left(t_{1}\right)/\hbar}\right)\right\}
\]

Using cumulant expansion and writing $Q_{j}(t)$ in terms of creation
and annihilation operators, the above expression can be evaluated.
Then we get
\[
Tr_{ph}\left\{ \left(\hat{T}e^{-i\int_{0}^{t}dt_{1}\varepsilon_{n}(\mathbf{Q}(t_{1}))/\hbar}\right)\rho_{ph}(0)\left(\hat{T}^{\dagger}\, e^{i\int_{0}^{t}dt_{1}\varepsilon_{n'}(\mathbf{Q}(t_{1}))/\hbar}\right)\right\} =e^{-i\varepsilon_{nn'}t/\hbar}e^{-\phi_{n,n'}(t)}.
\]
where $ \varepsilon_{nn'}=\varepsilon_{n}^0-\varepsilon_{n'}^0$,
\begin{equation}
Re(\phi_{n,n'}(t))=\frac{1}{\hbar}\int_{0}^{\infty}d\omega J(\omega)\frac{1-\cos(\omega t)}{\omega^{2}}\coth\left(\frac{\beta\hbar\omega}{2}\right)\sum_{j=1,2,...,7}\left(\frac{\partial\varepsilon_{n}}{\partial Q_{j}}-\frac{\partial\varepsilon_{n'}}{\partial Q_{j}}\right)_{Q_{j}=0}^{2}\label{eq:Rephi-1}
\end{equation}
and
\begin{equation}
Im(\phi_{n,n'}(t))=\frac{1}{\hbar}\int_{0}^{\infty}d\omega J(\omega)\frac{\sin(\omega t)-\omega t}{\omega^{2}}\sum_{j=1,2,...,7}\left(\left(\frac{\partial\varepsilon_{n}}{\partial Q_{j}}\right)_{Q_{j}=0}^{2}-\left(\frac{\partial\varepsilon_{n'}}{\partial Q_{j}}\right)_{Q_{j}=0}^{2}\right)\label{eq:Imphi-1}
\end{equation}

We now evaluate the two integrals. First in Eq. (\ref{eq:Rephi}),
we rewrite
\begin{eqnarray}
\frac{1}{\hbar}\int_{0}^{\infty}d\omega J(\omega)\frac{1-\cos(\omega t)}{\omega^{2}}\coth\left(\frac{\beta\hbar\omega}{2}\right) & = & \frac{1}{2\hbar}\int_{0}^{t}dt_{2}\int_{0}^{t_{2}}dt_{1}\int_{-\infty}^{\infty}d\omega J(\omega)e^{i\omega t_{1}}\coth\left(\frac{\beta\hbar\omega}{2}\right).
\end{eqnarray}
Using Drude spectral density, $\int_{-\infty}^{\infty}d\omega J(\omega)e^{i\omega t_{1}}\coth\left(\frac{\beta\hbar\omega}{2}\right)=\frac{2\lambda}{\pi}\int_{-\infty}^{\infty}d\omega\frac{\omega}{\omega^{2}+\omega_{C}^{2}}e^{i\omega t_{1}}\coth\left(\frac{\beta\hbar\omega}{2}\right)$
which can be evaluated using contour integration. The above function
has poles at $\omega=\pm i\pi,\pm i\frac{2\pi n}{\beta\hbar}\,(n=0,1,2,...)$.
Applying the Cauchy Integral theorem at the poles and evaluating the
time-integration, we have
\begin{eqnarray}
\frac{1}{\hbar}\int_{0}^{\infty}d\omega J(\omega)\frac{1-\cos(\omega t)}{\omega^{2}}\coth\left(\frac{\beta\hbar\omega}{2}\right)\nonumber \\
= & \frac{\lambda}{\hbar\omega_{c}}\,\cot\left(\frac{\beta\hbar\omega_{c}}{2}\right)\left(e^{-\omega_{c}t}+\omega_{c}t-1\right)\nonumber \\
 & +\frac{4\lambda\omega_{c}}{\hbar^{2}\beta}\sum_{n=1}^{\infty}\frac{1}{\nu_{n}\left(\nu_{n}^{2}-\omega_{c}^{2}\right)}\left(e^{-\nu_{n}t}+\nu_{n}t-1\right).
\end{eqnarray}
Here, $\nu_{n}=\frac{2\pi n}{\beta\hbar}$. Eq. (\ref{eq:Imphi})
could easily be evaluated to give the following expressions:
\begin{eqnarray}
\frac{1}{\hbar}\int_{0}^{\infty}d\omega J(\omega)\frac{\sin(\omega t)-\omega t}{\omega^{2}}=-\frac{\lambda}{\hbar\omega_{c}}\left(e^{-\omega_{c}t}+\omega_{c}t-1\right).
\end{eqnarray}
Therefore,
\begin{eqnarray}
Re(\phi_{n,n'}(t))=\left(\frac{\lambda}{\hbar\omega_{c}}\,\cot\left(\frac{\beta\hbar\omega_{c}}{2}\right)\left(e^{-\omega_{c}t}+\omega_{c}t-1\right)+\frac{4\lambda\omega_{c}}{\hbar^{2}\beta}\sum_{n=1}^{\infty}\frac{1}{\nu_{n}\left(\nu_{n}^{2}-\omega_{c}^{2}\right)}\left(e^{-\nu_{n}t}+\nu_{n}t-1\right)\right)\nonumber \\
\sum_{j=1,2,...,7}\left(\frac{\partial\varepsilon_{n}}{\partial Q_{j}}-\frac{\partial\varepsilon_{n'}}{\partial Q_{j}}\right)_{Q_{j}=0}^{2}\label{eq:rephi}
\end{eqnarray}
and
\begin{equation}
Im(\phi_{n,n'}(t))=\left(-\frac{\lambda}{\hbar\omega_{c}}\left(e^{-\omega_{c}t}+\omega_{c}t-1\right)\right)\sum_{j=1,2,...,7}\left(\left(\frac{\partial\varepsilon_{n}}{\partial Q_{j}}\right)_{Q_{j}=0}^{2}-\left(\frac{\partial\varepsilon_{n'}}{\partial Q_{j}}\right)_{Q_{j}=0}^{2}\right).\label{eq:imphi}
\end{equation}

\section{Deriving the Master Equation}
\label{AppendixB}
 In this section, we give a brief derivation of the master equation.  The approach is well known and is used in several papers \cite{Leegwater1997,KuehnSundstroem1997, Zhang:1998ri,Yang:2002ss,Renger&May1998,Renger&MarcusJPCA2003,RengerMarcusJCP2003}.  As is usual in deriving master equations
\cite{DiVentraBook}, we introduce two projection operators $\mathbb{A}$
and $\mathbb{B}$ which are defined by
\begin{equation}
\mathbb{A}X=\rho_{ph}\left(0\right)Tr_{ph}X\label{eq:ProjectionA}
\end{equation}
\begin{equation}
\mathbb{B}X=\left(1-\mathbb{A}\right)X\label{eq:Projection B}
\end{equation}
for any operator $X$. Using these and following \cite{DiVentraBook},
we derive the equations:
\begin{equation}
\frac{\partial\widetilde{\rho}_{1}(t)}{\partial t}=\mathbb{A}L_{na}(t)\widetilde{\rho}_{1}(t)+\mathbb{A}L_{na}(t)\widetilde{\rho}_{2}(t),\label{eq:rhotilde1}
\end{equation}
and
\begin{equation}
\frac{\partial\widetilde{\rho}_{2}(t)}{\partial t}=\mathbb{B}L_{na}(t)\widetilde{\rho}_{1}(t)+\mathbb{B}L_{na}(t)\widetilde{\rho}_{2}(t),\label{rhotilde2}
\end{equation}
where $\widetilde{\rho}_{1}(t)=\mathbb{A}\widetilde{\rho}_{  nn' }(t)$ and
$\widetilde{\rho}_{2}(t)=\mathbb{B}\widetilde{\rho}_{  nn' }(t)$.  Note that $\rho_1(t)$ and $\rho_2(t)$, are dependent on the initial coherence $|n\rangle \langle n'|$, which we have not indicated explicitly as that will clutter up the notation.   It
is easy to show that $\mathbb{A}L_{na}(t)\widetilde{\rho}_{1}(t)=0$.
Solving Eq. (\ref{rhotilde2}) and using it in Eq. (\ref{eq:rhotilde1})
we get
\begin{equation}
\frac{\partial\widetilde{\rho}_{1}(t)}{\partial t}=Tr_{ph}\left\{ L\left(t\right)\int_{0}^{t}dt_{1}e^{\mathbb{B}L_{na}\left(t-t_{1}\right)}L_{na}\left(t_{1}\right)\widetilde{\rho}_{ph}\left(0\right)\right\} \widetilde{\rho}_{1}(t_{1}).\label{eq:Lpho}
\end{equation}
As this is expected to be small, in Eq. (\ref{eq:Lpho}) we retain
terms upto second order in $L_{na}(t)$. This implies that we shall
approximate $e^{BL_{na}\left(t-t_{1}\right)}\approx1$. Further, we
make the Markovian approximation and put $\widetilde{\rho}_{1}(t_{1})\approx\widetilde{\rho}_{1}(t)$
and take the upper limit in the integral to be infinity, to get
\begin{equation}
\frac{\partial\widetilde{\rho}_{1}(t)}{\partial t}=Tr_{ph}\left\{ \int_{0}^{\infty}dt_{1}L_{na}\left(t\right)L_{na}\left(t_{1}\right)\widetilde{\rho}_{ph}\left(0\right)\right\} \widetilde{\rho}_{1}(t).\label{eq:equationforrho-1-Markov}
\end{equation}
On evaluating the right hand side of the above expression, we get
\begin{multline}
\frac{\partial\widetilde{\rho}_{1}(t)}{\partial t}=-\sum_{r,r',s}B_{r,r',s}[X_{r',s}|r\rangle\langle s|\widetilde{\rho}_{1}(t)+X_{r',r}\widetilde{\rho}_{1}(t)|r\rangle\langle s|]+\sum_{r,r',s',s}B'_{r,r',s',s}(X_{r,r'}+X_{s',s})[|r\rangle\langle r'|\widetilde{\rho}_{1}(t)|s'\rangle\langle s|],
\end{multline}
with the initial condition $\widetilde{\rho}_{1}(0)=|n\rangle\langle n'|$.
$|r\rangle$, $|r'\rangle$, $|s'\rangle$ and $|s\rangle$ are the
adiabatic states evaluated at $\mathbf{Q=0}.$ $B_{r,r',s}=\sum_{j}A_{r,r',\mathbf{0}}^{j}A_{r',s,\mathbf{0}}^{j}$;
$B'_{r,r',s',s}=\sum_{j}A_{r,r',\mathbf{0}}^{j}A_{s',s,\mathbf{0}}^{j}$
and $\omega_{rr'}=(\varepsilon_{r}^0-\varepsilon_{r'}^0)/\hbar$.
\begin{equation}
X_{r,r'}=\begin{cases}
\frac{\pi}{\hbar}\frac{J(\omega_{rr'})\omega_{rr'}^{2}}{e^{\beta\hbar\omega_{rr'}}-1} & \mbox{if }\omega_{rr'}>0\\
\frac{\pi}{\hbar}\frac{J(\omega_{r'r})\omega_{r'r}^{2}}{1-e^{-\beta\hbar\omega_{r'r}}} & \mbox{if }\omega_{r'r}>0.
\end{cases}
\end{equation}
Here $J_{j}(\omega)$ is the spectral density defined as
\begin{equation}
J_{j}(\omega)=\sum_{k}\frac{m_{jk}\nu_{jk}^{2}}{2\omega_{jk}}\delta(\omega-\omega_{jk}).
\end{equation}


\begin{thebibliography}{47}%
\makeatletter
\providecommand \@ifxundefined [1]{%
 \@ifx{#1\undefined}
}%
\providecommand \@ifnum [1]{%
 \ifnum #1\expandafter \@firstoftwo
 \else \expandafter \@secondoftwo
 \fi
}%
\providecommand \@ifx [1]{%
 \ifx #1\expandafter \@firstoftwo
 \else \expandafter \@secondoftwo
 \fi
}%
\providecommand \natexlab [1]{#1}%
\providecommand \enquote  [1]{``#1''}%
\providecommand \bibnamefont  [1]{#1}%
\providecommand \bibfnamefont [1]{#1}%
\providecommand \citenamefont [1]{#1}%
\providecommand \href@noop [0]{\@secondoftwo}%
\providecommand \href [0]{\begingroup \@sanitize@url \@href}%
\providecommand \@href[1]{\@@startlink{#1}\@@href}%
\providecommand \@@href[1]{\endgroup#1\@@endlink}%
\providecommand \@sanitize@url [0]{\catcode `\\12\catcode `\$12\catcode
  `\&12\catcode `\#12\catcode `\^12\catcode `\_12\catcode `\%12\relax}%
\providecommand \@@startlink[1]{}%
\providecommand \@@endlink[0]{}%
\providecommand \url  [0]{\begingroup\@sanitize@url \@url }%
\providecommand \@url [1]{\endgroup\@href {#1}{\urlprefix }}%
\providecommand \urlprefix  [0]{URL }%
\providecommand \Eprint [0]{\href }%
\@ifxundefined \urlstyle {%
  \providecommand \doi  [0]{\begingroup \@sanitize@url \@doi}%
  \providecommand \@doi [1]{\endgroup \@@startlink {\doibase
  #1}doi:\discretionary {}{}{}#1\@@endlink }%
}{%
  \providecommand \doi  [0]{doi:\discretionary{}{}{}\begingroup
  \urlstyle{rm}\Url }%
}%
\providecommand \doibase [0]{http://dx.doi.org/}%
\providecommand \Doi [0]{\begingroup \@sanitize@url \@Doi }%
\providecommand \@Doi  [1]{\endgroup\@@startlink{\doibase#1}\@@Doi}%
\providecommand \@@Doi [1]{#1\@@endlink}%
\providecommand \selectlanguage [0]{\@gobble}%
\providecommand \bibinfo  [0]{\@secondoftwo}%
\providecommand \bibfield  [0]{\@secondoftwo}%
\providecommand \translation [1]{[#1]}%
\providecommand \BibitemOpen [0]{}%
\providecommand \bibitemStop [0]{}%
\providecommand \bibitemNoStop [0]{.\EOS\space}%
\providecommand \EOS [0]{\spacefactor3000\relax}%
\providecommand \BibitemShut  [1]{\csname bibitem#1\endcsname}%
%</preamble>
\bibitem [{\citenamefont {Brixner}\ \emph {et~al.}(2005)\citenamefont
  {Brixner}, \citenamefont {Stenger}, \citenamefont {Vaswani}, \citenamefont
  {Cho}, \citenamefont {Blankenship},\ and\ \citenamefont
  {Fleming}}]{Brixner:2005rz}%
  \BibitemOpen
  \bibfield  {author} {\bibinfo {author} {\bibfnamefont {T.}~\bibnamefont
  {Brixner}}, \bibinfo {author} {\bibfnamefont {J.}~\bibnamefont {Stenger}},
  \bibinfo {author} {\bibfnamefont {H.}~\bibnamefont {Vaswani}}, \bibinfo
  {author} {\bibfnamefont {M.}~\bibnamefont {Cho}}, \bibinfo {author}
  {\bibfnamefont {R.}~\bibnamefont {Blankenship}}, \ and\ \bibinfo {author}
  {\bibfnamefont {G.}~\bibnamefont {Fleming}},\ }\Doi {10.1038/nature03429}
  {\bibfield  {journal} {\bibinfo  {journal} {Nature},\ }\textbf {\bibinfo
  {volume} {434}},\ \bibinfo {pages} {625} (\bibinfo {year}
  {2005})}\BibitemShut {NoStop}%
\bibitem [{\citenamefont {Engel}\ \emph {et~al.}(2007)\citenamefont {Engel},
  \citenamefont {Calhoun}, \citenamefont {Read}, \citenamefont {Ahn},
  \citenamefont {Mancal}, \citenamefont {Cheng}, \citenamefont {Blankenship},\
  and\ \citenamefont {Fleming}}]{Engel:2007od}%
  \BibitemOpen
  \bibfield  {author} {\bibinfo {author} {\bibfnamefont {G.}~\bibnamefont
  {Engel}}, \bibinfo {author} {\bibfnamefont {T.}~\bibnamefont {Calhoun}},
  \bibinfo {author} {\bibfnamefont {E.}~\bibnamefont {Read}}, \bibinfo {author}
  {\bibfnamefont {T.-K.}\ \bibnamefont {Ahn}}, \bibinfo {author} {\bibfnamefont
  {T.}~\bibnamefont {Mancal}}, \bibinfo {author} {\bibfnamefont {Y.-C.}\
  \bibnamefont {Cheng}}, \bibinfo {author} {\bibfnamefont {R.}~\bibnamefont
  {Blankenship}}, \ and\ \bibinfo {author} {\bibfnamefont {G.}~\bibnamefont
  {Fleming}},\ }\Doi {10.1038/nature05678} {\bibfield  {journal} {\bibinfo
  {journal} {Nature},\ }\textbf {\bibinfo {volume} {446}},\ \bibinfo {pages}
  {782} (\bibinfo {year} {2007})}\BibitemShut {NoStop}%
\bibitem [{\citenamefont {Read}\ \emph {et~al.}(2007)\citenamefont {Read},
  \citenamefont {Engel}, \citenamefont {Calhoun}, \citenamefont {Mancal},
  \citenamefont {Ahn}, \citenamefont {Blankenship},\ and\ \citenamefont
  {Fleming}}]{Read:2007kt}%
  \BibitemOpen
  \bibfield  {author} {\bibinfo {author} {\bibfnamefont {E.}~\bibnamefont
  {Read}}, \bibinfo {author} {\bibfnamefont {G.}~\bibnamefont {Engel}},
  \bibinfo {author} {\bibfnamefont {T.}~\bibnamefont {Calhoun}}, \bibinfo
  {author} {\bibfnamefont {T.}~\bibnamefont {Mancal}}, \bibinfo {author}
  {\bibfnamefont {T.}~\bibnamefont {Ahn}}, \bibinfo {author} {\bibfnamefont
  {R.}~\bibnamefont {Blankenship}}, \ and\ \bibinfo {author} {\bibfnamefont
  {G.}~\bibnamefont {Fleming}},\ }\Doi {10.1073/pnas.0701201104} {\bibfield
  {journal} {\bibinfo  {journal} {Proceedings of the National Academy of
  Sciences of the United States of America},\ }\textbf {\bibinfo {volume}
  {104}},\ \bibinfo {pages} {14203} (\bibinfo {year} {2007})}\BibitemShut
  {NoStop}%
\bibitem [{\citenamefont {Leegwater}\ \emph {et~al.}(1997)\citenamefont
  {Leegwater}, \citenamefont {Durrant},\ and\ \citenamefont
  {Klug}}]{Leegwater1997}%
  \BibitemOpen
  \bibfield  {author} {\bibinfo {author} {\bibfnamefont {J.~A.}\ \bibnamefont
  {Leegwater}}, \bibinfo {author} {\bibfnamefont {J.~R.}\ \bibnamefont
  {Durrant}}, \ and\ \bibinfo {author} {\bibfnamefont {D.~R.}\ \bibnamefont
  {Klug}},\ }\Doi {10.1021/jp9634058} {\bibfield  {journal} {\bibinfo
  {journal} {The Journal of Physical Chemistry B},\ }\textbf {\bibinfo {volume}
  {101}},\ \bibinfo {pages} {7205} (\bibinfo {year} {1997})},\ \Eprint
  {http://arxiv.org/abs/http://pubs.acs.org/doi/pdf/10.1021/jp9634058}
  {http://pubs.acs.org/doi/pdf/10.1021/jp9634058} \BibitemShut {NoStop}%
\bibitem [{\citenamefont {Kuehn}\ and\ \citenamefont
  {Sundstroem}(1997)}]{KuehnSundstroem1997}%
  \BibitemOpen
  \bibfield  {author} {\bibinfo {author} {\bibfnamefont {O.}~\bibnamefont
  {Kuehn}}\ and\ \bibinfo {author} {\bibfnamefont {V.}~\bibnamefont
  {Sundstroem}},\ }\href@noop {} {\bibfield  {journal} {\bibinfo  {journal}
  {Journal of Chemical Physics},\ }\textbf {\bibinfo {volume} {107}},\ \bibinfo
  {pages} {4154} (\bibinfo {year} {1997})}\BibitemShut {NoStop}%
\bibitem [{\citenamefont {Renger}\ and\ \citenamefont
  {May}(1998)}]{Renger&May1998}%
  \BibitemOpen
  \bibfield  {author} {\bibinfo {author} {\bibfnamefont {T.}~\bibnamefont
  {Renger}}\ and\ \bibinfo {author} {\bibfnamefont {V.}~\bibnamefont {May}},\
  }\Doi {10.1021/jp9800665} {\bibfield  {journal} {\bibinfo  {journal} {The
  Journal of Physical Chemistry A},\ }\textbf {\bibinfo {volume} {102}},\
  \bibinfo {pages} {4381} (\bibinfo {year} {1998})},\ \Eprint
  {http://arxiv.org/abs/http://pubs.acs.org/doi/pdf/10.1021/jp9800665}
  {http://pubs.acs.org/doi/pdf/10.1021/jp9800665} \BibitemShut {NoStop}%
\bibitem [{\citenamefont {Fenna}\ and\ \citenamefont
  {Matthews}(1975)}]{Fenna:1975}%
  \BibitemOpen
  \bibfield  {author} {\bibinfo {author} {\bibfnamefont {R.}~\bibnamefont
  {Fenna}}\ and\ \bibinfo {author} {\bibfnamefont {B.}~\bibnamefont
  {Matthews}},\ }\href@noop {} {\bibfield  {journal} {\bibinfo  {journal}
  {Nature},\ }\textbf {\bibinfo {volume} {258}},\ \bibinfo {pages} {573}
  (\bibinfo {year} {1975})}\BibitemShut {NoStop}%
\bibitem [{\citenamefont {Li}\ \emph {et~al.}(1997)\citenamefont {Li},
  \citenamefont {Zhou}, \citenamefont {Blankenship},\ and\ \citenamefont
  {Allen}}]{Blankenship:1997}%
  \BibitemOpen
  \bibfield  {author} {\bibinfo {author} {\bibfnamefont {Y.-F.}\ \bibnamefont
  {Li}}, \bibinfo {author} {\bibfnamefont {W.}~\bibnamefont {Zhou}}, \bibinfo
  {author} {\bibfnamefont {R.}~\bibnamefont {Blankenship}}, \ and\ \bibinfo
  {author} {\bibfnamefont {J.}~\bibnamefont {Allen}},\ }\href@noop {}
  {\bibfield  {journal} {\bibinfo  {journal} {The Journal of Molecular
  Biology},\ }\textbf {\bibinfo {volume} {271}},\ \bibinfo {pages} {456}
  (\bibinfo {year} {1997})}\BibitemShut {NoStop}%
\bibitem [{\citenamefont {Camara-Artigas}\ \emph {et~al.}(2003)\citenamefont
  {Camara-Artigas}, \citenamefont {Blankenship},\ and\ \citenamefont
  {Allen}}]{Blankenship:2003}%
  \BibitemOpen
  \bibfield  {author} {\bibinfo {author} {\bibfnamefont {A.}~\bibnamefont
  {Camara-Artigas}}, \bibinfo {author} {\bibfnamefont {R.}~\bibnamefont
  {Blankenship}}, \ and\ \bibinfo {author} {\bibfnamefont {J.}~\bibnamefont
  {Allen}},\ }\href@noop {} {\bibfield  {journal} {\bibinfo  {journal}
  {Photosynthesis Research},\ }\textbf {\bibinfo {volume} {75}},\ \bibinfo
  {pages} {49} (\bibinfo {year} {2003})}\BibitemShut {NoStop}%
\bibitem [{\citenamefont {Panitchayangkoon}\ \emph {et~al.}(2010)\citenamefont
  {Panitchayangkoon}, \citenamefont {Hayes}, \citenamefont {Fransted},
  \citenamefont {Caram}, \citenamefont {Harel}, \citenamefont {Wen},
  \citenamefont {Blankenship},\ and\ \citenamefont
  {Engel}}]{Panitchayangkoon:2010lo}%
  \BibitemOpen
  \bibfield  {author} {\bibinfo {author} {\bibfnamefont {G.}~\bibnamefont
  {Panitchayangkoon}}, \bibinfo {author} {\bibfnamefont {D.}~\bibnamefont
  {Hayes}}, \bibinfo {author} {\bibfnamefont {K.}~\bibnamefont {Fransted}},
  \bibinfo {author} {\bibfnamefont {J.}~\bibnamefont {Caram}}, \bibinfo
  {author} {\bibfnamefont {E.}~\bibnamefont {Harel}}, \bibinfo {author}
  {\bibfnamefont {J.}~\bibnamefont {Wen}}, \bibinfo {author} {\bibfnamefont
  {R.}~\bibnamefont {Blankenship}}, \ and\ \bibinfo {author} {\bibfnamefont
  {G.}~\bibnamefont {Engel}},\ }\Doi {10.1073/pnas.1005484107} {\bibfield
  {journal} {\bibinfo  {journal} {Proceedings of the National Academy of
  Sciences of the United States of America},\ }\textbf {\bibinfo {volume}
  {107}},\ \bibinfo {pages} {12766} (\bibinfo {year} {2010})}\BibitemShut
  {NoStop}%
\bibitem [{\citenamefont {Prezhdo}\ and\ \citenamefont
  {Rossky}(1998)}]{Prezhdo:1998ij}%
  \BibitemOpen
  \bibfield  {author} {\bibinfo {author} {\bibfnamefont {O.}~\bibnamefont
  {Prezhdo}}\ and\ \bibinfo {author} {\bibfnamefont {P.~J.}\ \bibnamefont
  {Rossky}},\ }\href@noop {} {\bibfield  {journal} {\bibinfo  {journal}
  {Physical Review Letters},\ }\textbf {\bibinfo {volume} {81}},\ \bibinfo
  {pages} {5294} (\bibinfo {year} {1998})}\BibitemShut {NoStop}%
\bibitem [{\citenamefont {Gilmore}\ and\ \citenamefont
  {McKenzie}(2006)}]{Gilmore:2006vy}%
  \BibitemOpen
  \bibfield  {author} {\bibinfo {author} {\bibfnamefont {J.}~\bibnamefont
  {Gilmore}}\ and\ \bibinfo {author} {\bibfnamefont {R.}~\bibnamefont
  {McKenzie}},\ }\Doi {10.1016/j.cplett.2005.12.104} {\bibfield  {journal}
  {\bibinfo  {journal} {Chemical physics letters}} (\bibinfo {year} {2006})},\
  \doi {10.1016/j.cplett.2005.12.104}\BibitemShut {NoStop}%
\bibitem [{\citenamefont {Gilmore}\ and\ \citenamefont
  {McKenzie}(2008)}]{Gilmore:JPhysChemA2008}%
  \BibitemOpen
  \bibfield  {author} {\bibinfo {author} {\bibfnamefont {J.}~\bibnamefont
  {Gilmore}}\ and\ \bibinfo {author} {\bibfnamefont {R.~H.}\ \bibnamefont
  {McKenzie}},\ }\href@noop {} {\bibfield  {journal} {\bibinfo  {journal}
  {Journal of Physical Chemistry A},\ }\textbf {\bibinfo {volume} {112}},\
  \bibinfo {pages} {2162} (\bibinfo {year} {2008})}\BibitemShut {NoStop}%
\bibitem [{\citenamefont {Nalbach}\ \emph
  {et~al.}(2011){\natexlab{a}}\citenamefont {Nalbach}, \citenamefont
  {Ishizaki}, \citenamefont {Fleming},\ and\ \citenamefont
  {Thorwart}}]{Nalbach:2011gr}%
  \BibitemOpen
  \bibfield  {author} {\bibinfo {author} {\bibfnamefont {P.}~\bibnamefont
  {Nalbach}}, \bibinfo {author} {\bibfnamefont {A.}~\bibnamefont {Ishizaki}},
  \bibinfo {author} {\bibfnamefont {G.~R.}\ \bibnamefont {Fleming}}, \ and\
  \bibinfo {author} {\bibfnamefont {M.}~\bibnamefont {Thorwart}},\ }\Doi
  {10.1088/1367-2630/13/6/063040} {\bibfield  {journal} {\bibinfo  {journal}
  {New Journal of Physics},\ }\textbf {\bibinfo {volume} {13}} (\bibinfo {year}
  {2011}{\natexlab{a}})},\ \doi {10.1088/1367-2630/13/6/063040}\BibitemShut
  {NoStop}%
\bibitem [{\citenamefont {Rebentrost}\ \emph
  {et~al.}(2009){\natexlab{a}}\citenamefont {Rebentrost}, \citenamefont
  {Mohseni},\ and\ \citenamefont {Aspuru-Guzik}}]{Rebentrost:2009ve}%
  \BibitemOpen
  \bibfield  {author} {\bibinfo {author} {\bibfnamefont {P.}~\bibnamefont
  {Rebentrost}}, \bibinfo {author} {\bibfnamefont {M.}~\bibnamefont {Mohseni}},
  \ and\ \bibinfo {author} {\bibfnamefont {A.}~\bibnamefont {Aspuru-Guzik}},\
  }\Doi {10.1021/jp901724d} {\bibfield  {journal} {\bibinfo  {journal} {The
  Journal of Physical Chemistry. B},\ }\textbf {\bibinfo {volume} {113}},\
  \bibinfo {pages} {9942} (\bibinfo {year} {2009}{\natexlab{a}})}\BibitemShut
  {NoStop}%
\bibitem [{\citenamefont {Rebentrost}\ \emph
  {et~al.}(2009){\natexlab{b}}\citenamefont {Rebentrost}, \citenamefont
  {Mohseni}, \citenamefont {Kassal}, \citenamefont {Lloyd},\ and\ \citenamefont
  {Aspuru-Guzik}}]{Rebentrost:2009kls}%
  \BibitemOpen
  \bibfield  {author} {\bibinfo {author} {\bibfnamefont {P.}~\bibnamefont
  {Rebentrost}}, \bibinfo {author} {\bibfnamefont {M.}~\bibnamefont {Mohseni}},
  \bibinfo {author} {\bibfnamefont {I.}~\bibnamefont {Kassal}}, \bibinfo
  {author} {\bibfnamefont {S.}~\bibnamefont {Lloyd}}, \ and\ \bibinfo {author}
  {\bibfnamefont {A.}~\bibnamefont {Aspuru-Guzik}},\ }\href@noop {} {\bibfield
  {journal} {\bibinfo  {journal} {New Journal of Physics},\ }\textbf {\bibinfo
  {volume} {11}},\ \bibinfo {pages} {033003} (\bibinfo {year}
  {2009}{\natexlab{b}})}\BibitemShut {NoStop}%
\bibitem [{\citenamefont {Abramavicius}\ and\ \citenamefont
  {Mukamel}(2010)}]{Abramavicius:2010qu}%
  \BibitemOpen
  \bibfield  {author} {\bibinfo {author} {\bibfnamefont {D.}~\bibnamefont
  {Abramavicius}}\ and\ \bibinfo {author} {\bibfnamefont {S.}~\bibnamefont
  {Mukamel}},\ }\Doi {10.1063/1.3493580} {\bibfield  {journal} {\bibinfo
  {journal} {The Journal of chemical physics},\ }\textbf {\bibinfo {volume}
  {133}} (\bibinfo {year} {2010})},\ \doi {10.1063/1.3493580}\BibitemShut
  {NoStop}%
\bibitem [{\citenamefont {Singh}(2012)}]{Singh:condmat2012}%
  \BibitemOpen
  \bibfield  {author} {\bibinfo {author} {\bibfnamefont {N.}~\bibnamefont
  {Singh}},\ }\href@noop {} {\bibfield  {journal} {\bibinfo  {journal}
  {arXiv:1203.0147v1 [cond-mat.stat-mech]}} (\bibinfo {year}
  {2012})}\BibitemShut {NoStop}%
\bibitem [{\citenamefont {Ball}(2011)}]{Ball:2011nl}%
  \BibitemOpen
  \bibfield  {author} {\bibinfo {author} {\bibfnamefont {P.}~\bibnamefont
  {Ball}},\ }\Doi {10.1038/474272a} {\bibfield  {journal} {\bibinfo  {journal}
  {Nature},\ }\textbf {\bibinfo {volume} {474}},\ \bibinfo {pages} {272}
  (\bibinfo {year} {2011})}\BibitemShut {NoStop}%
\bibitem [{\citenamefont {Lloyd}(2011)}]{Lloyd:2011dl}%
  \BibitemOpen
  \bibfield  {author} {\bibinfo {author} {\bibfnamefont {S.}~\bibnamefont
  {Lloyd}},\ }\Doi {10.1088/1742-6596/302/1/012037} {\bibfield  {journal}
  {\bibinfo  {journal} {Journal of Physics: Conference Series},\ }\textbf
  {\bibinfo {volume} {302}} (\bibinfo {year} {2011})},\ \doi
  {10.1088/1742-6596/302/1/012037}\BibitemShut {NoStop}%
\bibitem [{\citenamefont {Ishizaki}\ and\ \citenamefont
  {Fleming}(2009){\natexlab{a}}}]{Ishizaki:2009sd}%
  \BibitemOpen
  \bibfield  {author} {\bibinfo {author} {\bibfnamefont {A.}~\bibnamefont
  {Ishizaki}}\ and\ \bibinfo {author} {\bibfnamefont {G.}~\bibnamefont
  {Fleming}},\ }\Doi {10.1073/pnas.0908989106} {\bibfield  {journal} {\bibinfo
  {journal} {Proceedings of the National Academy of Sciences of the United
  States of America},\ }\textbf {\bibinfo {volume} {106}},\ \bibinfo {pages}
  {17255} (\bibinfo {year} {2009}{\natexlab{a}})}\BibitemShut {NoStop}%
\bibitem [{\citenamefont {Ishizaki}\ and\ \citenamefont
  {Fleming}(2009){\natexlab{b}}}]{Ishizaki:2009zm}%
  \BibitemOpen
  \bibfield  {author} {\bibinfo {author} {\bibfnamefont {A.}~\bibnamefont
  {Ishizaki}}\ and\ \bibinfo {author} {\bibfnamefont {G.}~\bibnamefont
  {Fleming}},\ }\Doi {10.1063/1.3155372} {\bibfield  {journal} {\bibinfo
  {journal} {The Journal of Chemical Physics},\ }\textbf {\bibinfo {volume}
  {130}},\ \bibinfo {pages} {23411} (\bibinfo {year}
  {2009}{\natexlab{b}})}\BibitemShut {NoStop}%
\bibitem [{\citenamefont {Mohseni}\ \emph {et~al.}(2008)\citenamefont
  {Mohseni}, \citenamefont {Rebentrost}, \citenamefont {Lloyd},\ and\
  \citenamefont {Aspuru-Guzik}}]{Mohseni:2008wm}%
  \BibitemOpen
  \bibfield  {author} {\bibinfo {author} {\bibfnamefont {M.}~\bibnamefont
  {Mohseni}}, \bibinfo {author} {\bibfnamefont {P.}~\bibnamefont {Rebentrost}},
  \bibinfo {author} {\bibfnamefont {S.}~\bibnamefont {Lloyd}}, \ and\ \bibinfo
  {author} {\bibfnamefont {A.}~\bibnamefont {Aspuru-Guzik}},\ }\Doi
  {10.1063/1.3002335} {\bibfield  {journal} {\bibinfo  {journal} {The Journal
  of Chemical Physics},\ }\textbf {\bibinfo {volume} {129}} (\bibinfo {year}
  {2008})},\ \doi {10.1063/1.3002335}\BibitemShut {NoStop}%
\bibitem [{\citenamefont {Caruso}\ \emph {et~al.}(2009)\citenamefont {Caruso},
  \citenamefont {Chin}, \citenamefont {Datta}, \citenamefont {Huelga},\ and\
  \citenamefont {Plenio}}]{Caruso:2009pp}%
  \BibitemOpen
  \bibfield  {author} {\bibinfo {author} {\bibfnamefont {F.}~\bibnamefont
  {Caruso}}, \bibinfo {author} {\bibfnamefont {A.~W.}\ \bibnamefont {Chin}},
  \bibinfo {author} {\bibfnamefont {A.}~\bibnamefont {Datta}}, \bibinfo
  {author} {\bibfnamefont {S.~F.}\ \bibnamefont {Huelga}}, \ and\ \bibinfo
  {author} {\bibfnamefont {M.~B.}\ \bibnamefont {Plenio}},\ }\Doi
  {10.1063/1.3223548} {\bibfield  {journal} {\bibinfo  {journal} {The Journal
  of Chemical Physics},\ }\textbf {\bibinfo {volume} {131}} (\bibinfo {year}
  {2009})},\ \doi {10.1063/1.3223548}\BibitemShut {NoStop}%
\bibitem [{\citenamefont {Olaya-Castro}\ and\ \citenamefont
  {Scholes}(2011)}]{Olaya-Castro:2011zn}%
  \BibitemOpen
  \bibfield  {author} {\bibinfo {author} {\bibfnamefont {A.}~\bibnamefont
  {Olaya-Castro}}\ and\ \bibinfo {author} {\bibfnamefont {G.}~\bibnamefont
  {Scholes}},\ }\Doi {10.1080/0144235X.2010.537060} {\bibfield  {journal}
  {\bibinfo  {journal} {International Reviews in Physical Chemistry},\ }\textbf
  {\bibinfo {volume} {30}},\ \bibinfo {pages} {49} (\bibinfo {year}
  {2011})}\BibitemShut {NoStop}%
\bibitem [{\citenamefont {Nalbach}\ \emph
  {et~al.}(2011){\natexlab{b}}\citenamefont {Nalbach}, \citenamefont {Braun},\
  and\ \citenamefont {Thorwart}}]{Nalbach:2011}%
  \BibitemOpen
  \bibfield  {author} {\bibinfo {author} {\bibfnamefont {P.}~\bibnamefont
  {Nalbach}}, \bibinfo {author} {\bibfnamefont {D.}~\bibnamefont {Braun}}, \
  and\ \bibinfo {author} {\bibfnamefont {M.}~\bibnamefont {Thorwart}},\
  }\href@noop {} {\bibfield  {journal} {\bibinfo  {journal} {Physical Review
  E},\ }\textbf {\bibinfo {volume} {84}},\ \bibinfo {pages} {041926} (\bibinfo
  {year} {2011}{\natexlab{b}})}\BibitemShut {NoStop}%
\bibitem [{\citenamefont {Olbrich}\ \emph
  {et~al.}(2011){\natexlab{a}}\citenamefont {Olbrich}, \citenamefont
  {Str{\"u}mpfer}, \citenamefont {Schulten},\ and\ \citenamefont
  {Kleinekathofer}}]{olbrichtheory2011}%
  \BibitemOpen
  \bibfield  {author} {\bibinfo {author} {\bibfnamefont {C.}~\bibnamefont
  {Olbrich}}, \bibinfo {author} {\bibfnamefont {J.}~\bibnamefont
  {Str{\"u}mpfer}}, \bibinfo {author} {\bibfnamefont {K.}~\bibnamefont
  {Schulten}}, \ and\ \bibinfo {author} {\bibfnamefont {U.}~\bibnamefont
  {Kleinekathofer}},\ }\Doi {10.1021/jz2007676} {\bibfield  {journal} {\bibinfo
   {journal} {The Journal of Physical Chemistry Letters},\ }\textbf {\bibinfo
  {volume} {2011}},\ \bibinfo {pages} {1771} (\bibinfo {year}
  {2011}{\natexlab{a}})}\BibitemShut {NoStop}%
\bibitem [{\citenamefont {M$\ddot{u}$hlbacher}\ and\ \citenamefont
  {Kleinekathofer}(2012)}]{Kleinekathofer2012}%
  \BibitemOpen
  \bibfield  {author} {\bibinfo {author} {\bibfnamefont {L.}~\bibnamefont
  {M$\ddot{u}$hlbacher}}\ and\ \bibinfo {author} {\bibfnamefont
  {U.}~\bibnamefont {Kleinekathofer}},\ }\href@noop {} {\bibfield  {journal}
  {\bibinfo  {journal} {The Journal of Physical Chemistry B},\ }\textbf
  {\bibinfo {volume} {116}},\ \bibinfo {pages} {3900} (\bibinfo {year}
  {2012})}\BibitemShut {NoStop}%
\bibitem [{\citenamefont {Weiss}(2008)}]{WeissBook}%
  \BibitemOpen
  \bibfield  {author} {\bibinfo {author} {\bibfnamefont {U.}~\bibnamefont
  {Weiss}},\ }\href@noop {} {\emph {\bibinfo {title} {Quantum Dissipative
  Systems}}},\ \bibinfo {edition} {3rd}\ ed.,\ Series in Modern Condensed
  Matter Physics\ (\bibinfo  {publisher} {World Scientific},\ \bibinfo
  {address} {Singapore},\ \bibinfo {year} {2008})\BibitemShut {NoStop}%
\bibitem [{\citenamefont {Orth}\ \emph
  {et~al.}(2010){\natexlab{a}}\citenamefont {Orth}, \citenamefont {Imambekov},\
  and\ \citenamefont {Hur}}]{LeHur2010}%
  \BibitemOpen
  \bibfield  {author} {\bibinfo {author} {\bibfnamefont {P.~P.}\ \bibnamefont
  {Orth}}, \bibinfo {author} {\bibfnamefont {A.}~\bibnamefont {Imambekov}}, \
  and\ \bibinfo {author} {\bibfnamefont {K.~L.}\ \bibnamefont {Hur}},\
  }\href@noop {} {\bibfield  {journal} {\bibinfo  {journal} {Phys. Rev. A},\
  }\textbf {\bibinfo {volume} {82}},\ \bibinfo {pages} {032118} (\bibinfo
  {year} {2010}{\natexlab{a}})}\BibitemShut {NoStop}%
\bibitem [{\citenamefont {Orth}\ \emph
  {et~al.}(2010){\natexlab{b}}\citenamefont {Orth}, \citenamefont {Roosen},
  \citenamefont {Hofstetter},\ and\ \citenamefont {Hur}}]{LeHurPRB2010}%
  \BibitemOpen
  \bibfield  {author} {\bibinfo {author} {\bibfnamefont {P.~P.}\ \bibnamefont
  {Orth}}, \bibinfo {author} {\bibfnamefont {D.}~\bibnamefont {Roosen}},
  \bibinfo {author} {\bibfnamefont {W.}~\bibnamefont {Hofstetter}}, \ and\
  \bibinfo {author} {\bibfnamefont {K.~L.}\ \bibnamefont {Hur}},\ }\href@noop
  {} {\bibfield  {journal} {\bibinfo  {journal} {Phys. Rev. B},\ }\textbf
  {\bibinfo {volume} {82}},\ \bibinfo {pages} {144423} (\bibinfo {year}
  {2010}{\natexlab{b}})}\BibitemShut {NoStop}%
\bibitem [{\citenamefont {Schlosshauer}(2007)}]{SchlosshauerBook}%
  \BibitemOpen
  \bibfield  {author} {\bibinfo {author} {\bibfnamefont {M.}~\bibnamefont
  {Schlosshauer}},\ }\href@noop {} {\emph {\bibinfo {title} {Decoherence and
  the Quantum to Classical Transition}}}\ (\bibinfo  {publisher} {Springer},\
  \bibinfo {address} {Berlin},\ \bibinfo {year} {2007})\BibitemShut {NoStop}%
\bibitem [{\citenamefont {Sarovar}\ \emph {et~al.}(2011)\citenamefont
  {Sarovar}, \citenamefont {Cheng},\ and\ \citenamefont
  {Whaley}}]{Sarovar:2011}%
  \BibitemOpen
  \bibfield  {author} {\bibinfo {author} {\bibfnamefont {M.}~\bibnamefont
  {Sarovar}}, \bibinfo {author} {\bibfnamefont {Y.}~\bibnamefont {Cheng}}, \
  and\ \bibinfo {author} {\bibfnamefont {K.}~\bibnamefont {Whaley}},\
  }\href@noop {} {\bibfield  {journal} {\bibinfo  {journal} {Physical Review
  E},\ }\textbf {\bibinfo {volume} {83}},\ \bibinfo {pages} {011906} (\bibinfo
  {year} {2011})}\BibitemShut {NoStop}%
\bibitem [{\citenamefont {Ishizaki}\ and\ \citenamefont
  {Fleming}(2009){\natexlab{c}}}]{Ishizaki:2009mw}%
  \BibitemOpen
  \bibfield  {author} {\bibinfo {author} {\bibfnamefont {A.}~\bibnamefont
  {Ishizaki}}\ and\ \bibinfo {author} {\bibfnamefont {G.}~\bibnamefont
  {Fleming}},\ }\Doi {10.1063/1.3155214} {\bibfield  {journal} {\bibinfo
  {journal} {The Journal of Chemical Physics},\ }\textbf {\bibinfo {volume}
  {130}},\ \bibinfo {pages} {23411} (\bibinfo {year}
  {2009}{\natexlab{c}})}\BibitemShut {NoStop}%
\bibitem [{\citenamefont {Baer}(2006)}]{Baer-book}%
  \BibitemOpen
  \bibfield  {author} {\bibinfo {author} {\bibfnamefont {M.}~\bibnamefont
  {Baer}},\ }\href@noop {} {\emph {\bibinfo {title} {Beyond Born-Oppenheimer.
  Conical Intersections and Electronic Nonadiabatic Coupling Terms}}}\
  (\bibinfo  {publisher} {Wiley-Interscinece, New York},\ \bibinfo {year}
  {2006})\BibitemShut {NoStop}%
\bibitem [{\citenamefont {Zhang}\ \emph {et~al.}(1998)\citenamefont {Zhang},
  \citenamefont {Meier}, \citenamefont {Chernyak},\ and\ \citenamefont
  {Mukamel}}]{Zhang:1998ri}%
  \BibitemOpen
  \bibfield  {author} {\bibinfo {author} {\bibfnamefont {W.~M.}\ \bibnamefont
  {Zhang}}, \bibinfo {author} {\bibfnamefont {T.}~\bibnamefont {Meier}},
  \bibinfo {author} {\bibfnamefont {V.}~\bibnamefont {Chernyak}}, \ and\
  \bibinfo {author} {\bibfnamefont {S.}~\bibnamefont {Mukamel}},\ }\Doi
  {10.1063/1.476212} {\bibfield  {journal} {\bibinfo  {journal} {The Journal of
  Chemical Physics},\ }\textbf {\bibinfo {volume} {108}} (\bibinfo {year}
  {1998})},\ \doi {10.1063/1.476212}\BibitemShut {NoStop}%
\bibitem [{\citenamefont {Yang}\ and\ \citenamefont
  {Fleming}(2002)}]{Yang:2002ss}%
  \BibitemOpen
  \bibfield  {author} {\bibinfo {author} {\bibfnamefont {M.}~\bibnamefont
  {Yang}}\ and\ \bibinfo {author} {\bibfnamefont {G.}~\bibnamefont {Fleming}},\
  }\href@noop {} {\bibfield  {journal} {\bibinfo  {journal} {Chemical
  Physics},\ }\textbf {\bibinfo {volume} {282}},\ \bibinfo {pages} {163}
  (\bibinfo {year} {2002})}\BibitemShut {NoStop}%
\bibitem [{\citenamefont {Renger}\ and\ \citenamefont
  {Marcus}(2003){\natexlab{a}}}]{Renger&MarcusJPCA2003}%
  \BibitemOpen
  \bibfield  {author} {\bibinfo {author} {\bibfnamefont {T.}~\bibnamefont
  {Renger}}\ and\ \bibinfo {author} {\bibfnamefont {R.~A.}\ \bibnamefont
  {Marcus}},\ }\Doi {10.1021/jp026789c} {\bibfield  {journal} {\bibinfo
  {journal} {The Journal of Physical Chemistry A},\ }\textbf {\bibinfo {volume}
  {107}},\ \bibinfo {pages} {8404} (\bibinfo {year} {2003}{\natexlab{a}})},\
  \Eprint {http://arxiv.org/abs/http://pubs.acs.org/doi/pdf/10.1021/jp026789c}
  {http://pubs.acs.org/doi/pdf/10.1021/jp026789c} \BibitemShut {NoStop}%
\bibitem [{\citenamefont {Renger}\ and\ \citenamefont
  {Marcus}(2003){\natexlab{b}}}]{RengerMarcusJCP2003}%
  \BibitemOpen
  \bibfield  {author} {\bibinfo {author} {\bibfnamefont {T.}~\bibnamefont
  {Renger}}\ and\ \bibinfo {author} {\bibfnamefont {R.~A.}\ \bibnamefont
  {Marcus}},\ }\href@noop {} {\bibfield  {journal} {\bibinfo  {journal}
  {Journal of Chemical Physics},\ }\textbf {\bibinfo {volume} {116}},\ \bibinfo
  {pages} {9997} (\bibinfo {year} {2003}{\natexlab{b}})}\BibitemShut {NoStop}%
\bibitem [{\citenamefont {Adolphs}\ and\ \citenamefont
  {Renger}(2006)}]{AdolphsRenger:2006}%
  \BibitemOpen
  \bibfield  {author} {\bibinfo {author} {\bibfnamefont {J.}~\bibnamefont
  {Adolphs}}\ and\ \bibinfo {author} {\bibfnamefont {T.}~\bibnamefont
  {Renger}},\ }\href@noop {} {\bibfield  {journal} {\bibinfo  {journal}
  {Biophysical Journal},\ }\textbf {\bibinfo {volume} {91}},\ \bibinfo {pages}
  {2778} (\bibinfo {year} {2006})}\BibitemShut {NoStop}%
\bibitem [{\citenamefont {Nalbach}\ \emph {et~al.}(2010)\citenamefont
  {Nalbach}, \citenamefont {Eckel},\ and\ \citenamefont
  {Thorwart}}]{NalbachNJPCB2010}%
  \BibitemOpen
  \bibfield  {author} {\bibinfo {author} {\bibfnamefont {P.}~\bibnamefont
  {Nalbach}}, \bibinfo {author} {\bibfnamefont {J.}~\bibnamefont {Eckel}}, \
  and\ \bibinfo {author} {\bibfnamefont {M.}~\bibnamefont {Thorwart}},\
  }\href@noop {} {\bibfield  {journal} {\bibinfo  {journal} {New Journal of
  Physics},\ }\textbf {\bibinfo {volume} {12}},\ \bibinfo {pages} {065043}
  (\bibinfo {year} {2010})}\BibitemShut {NoStop}%
\bibitem [{\citenamefont {Olbrich}\ \emph
  {et~al.}(2011){\natexlab{b}}\citenamefont {Olbrich}, \citenamefont
  {Str{\"u}mpfer}, \citenamefont {Schulten},\ and\ \citenamefont
  {Kleinekathofer}}]{Olbrich:2011ez}%
  \BibitemOpen
  \bibfield  {author} {\bibinfo {author} {\bibfnamefont {C.}~\bibnamefont
  {Olbrich}}, \bibinfo {author} {\bibfnamefont {J.}~\bibnamefont
  {Str{\"u}mpfer}}, \bibinfo {author} {\bibfnamefont {K.}~\bibnamefont
  {Schulten}}, \ and\ \bibinfo {author} {\bibfnamefont {U.}~\bibnamefont
  {Kleinekathofer}},\ }\Doi {10.1021/jp1099514} {\bibfield  {journal} {\bibinfo
   {journal} {The Journal of Physical Chemistry B},\ }\textbf {\bibinfo
  {volume} {115}},\ \bibinfo {pages} {758} (\bibinfo {year}
  {2011}{\natexlab{b}})}\BibitemShut {NoStop}%
\bibitem [{\citenamefont {Chin}\ \emph {et~al.}(2013)\citenamefont {Chin},
  \citenamefont {Prior}, \citenamefont {Rosenbach}, \citenamefont
  {Caycedo-Soler}, \citenamefont {Huelga},\ and\ \citenamefont
  {Plenio}}]{PlenioNature2013}%
  \BibitemOpen
  \bibfield  {author} {\bibinfo {author} {\bibfnamefont {A.}~\bibnamefont
  {Chin}}, \bibinfo {author} {\bibfnamefont {J.}~\bibnamefont {Prior}},
  \bibinfo {author} {\bibfnamefont {R.}~\bibnamefont {Rosenbach}}, \bibinfo
  {author} {\bibfnamefont {F.}~\bibnamefont {Caycedo-Soler}}, \bibinfo {author}
  {\bibfnamefont {S.}~\bibnamefont {Huelga}}, \ and\ \bibinfo {author}
  {\bibfnamefont {M.}~\bibnamefont {Plenio}},\ }\href@noop {} {\bibfield
  {journal} {\bibinfo  {journal} {Nature Physics},\ }\textbf {\bibinfo {volume}
  {9}},\ \bibinfo {pages} {113} (\bibinfo {year} {2013})}\BibitemShut {NoStop}%
\bibitem [{\citenamefont {Tiwari}\ \emph {et~al.}(2013)\citenamefont {Tiwari},
  \citenamefont {Peters},\ and\ \citenamefont {Jonas}}]{TiwariPNAS2013}%
  \BibitemOpen
  \bibfield  {author} {\bibinfo {author} {\bibfnamefont {V.}~\bibnamefont
  {Tiwari}}, \bibinfo {author} {\bibfnamefont {W.~K.}\ \bibnamefont {Peters}},
  \ and\ \bibinfo {author} {\bibfnamefont {D.~M.}\ \bibnamefont {Jonas}},\
  }\href@noop {} {\bibfield  {journal} {\bibinfo  {journal} {Proceedings of the
  National Academy of Sciences},\ }\textbf {\bibinfo {volume} {110}},\ \bibinfo
  {pages} {1203} (\bibinfo {year} {2013})}\BibitemShut {NoStop}%
\bibitem [{\citenamefont {Polyutov}\ \emph {et~al.}(2012)\citenamefont
  {Polyutov}, \citenamefont {K{\"u}hn},\ and\ \citenamefont
  {Pullerits}}]{PolyutovCP2012}%
  \BibitemOpen
  \bibfield  {author} {\bibinfo {author} {\bibfnamefont {S.}~\bibnamefont
  {Polyutov}}, \bibinfo {author} {\bibfnamefont {O.}~\bibnamefont {K{\"u}hn}},
  \ and\ \bibinfo {author} {\bibfnamefont {T.}~\bibnamefont {Pullerits}},\
  }\href@noop {} {\bibfield  {journal} {\bibinfo  {journal} {Chemical
  Physics},\ }\textbf {\bibinfo {volume} {394}},\ \bibinfo {pages} {21}
  (\bibinfo {year} {2012})}\BibitemShut {NoStop}%
\bibitem [{\citenamefont {Christensson}\ \emph {et~al.}(2012)\citenamefont
  {Christensson}, \citenamefont {Kauffmann}, \citenamefont {Pullerits},\ and\
  \citenamefont {T.}}]{ChristenssonJPCB2012}%
  \BibitemOpen
  \bibfield  {author} {\bibinfo {author} {\bibfnamefont {N.}~\bibnamefont
  {Christensson}}, \bibinfo {author} {\bibfnamefont {H.}~\bibnamefont
  {Kauffmann}}, \bibinfo {author} {\bibfnamefont {T.}~\bibnamefont
  {Pullerits}}, \ and\ \bibinfo {author} {\bibfnamefont {M.}~\bibnamefont
  {T.}},\ }\href@noop {} {\bibfield  {journal} {\bibinfo  {journal} {The
  Journal of Physical Chemistry B},\ }\textbf {\bibinfo {volume} {116}},\
  \bibinfo {pages} {7449} (\bibinfo {year} {2012})}\BibitemShut {NoStop}%
\bibitem [{\citenamefont {DiVentra}(2008)}]{DiVentraBook}%
  \BibitemOpen
  \bibfield  {author} {\bibinfo {author} {\bibfnamefont {M.}~\bibnamefont
  {DiVentra}},\ }\href@noop {} {\emph {\bibinfo {title} {Electrical Transport
  in Nanoscale Systems}}}\ (\bibinfo  {publisher} {Cambridge University
  Press},\ \bibinfo {address} {New Delhi},\ \bibinfo {year} {2008})\BibitemShut
  {NoStop}%
\end{thebibliography}
\end{document}